\documentclass[epj]{svjour}
\usepackage{graphics}
\usepackage{epsfig}
\usepackage{amsmath}
\authorrunning{K.J. Pototzky et al.}
\titlerunning{Properties of odd nuclei and the impact of time-odd mean fields}
\begin{document}
\title{Properties of odd nuclei and the impact of time-odd mean fields: A systematic
Skyrme-Hartree-Fock analysis}
\author{K.J. Pototzky \inst{1} \and J. Erler \inst{1}
\and P.--G. Reinhard \inst{1} \and V.O. Nesterenko \inst{2}
}                     
%
%
\institute{Institut f\"ur Theoretische Physik II, Universit\"at
  Erlangen-N\"urnberg, Staudstr. 7,
D-91058 Erlangen, Germany \and Laboratory of Theoretical Physics, Joint Institute
for Nuclear Research, Dubna, Moscow region, 141980, Russia}
\date{Received: date / Revised version: date}
%
\abstract{
We present a systematic analysis of the description of odd nuclei
  by the Skyrme-Hartree-Fock approach augmented with pairing in BCS
  approximation and blocking of the odd nucleon.  Current and spin densities
  in the Skyrme functional produce time-odd mean fields (TOMF) for odd
  nuclei. Their effect on basic properties (binding energies, odd-even
  staggering, separation energies and spectra) is investigated for the three
  Skyrme parameterizations SkI3, SLy6, and SV-bas.  About 1300 spherical and
  axially-deformed odd nuclei with $16 \le Z \le 92$ are considered. The
  calculations demonstrate that the TOMF effect is generally small,
  although not fully negligible. The influence of the Skyrme parameterization
  and the consistency of the calculations are much more important. With a
  proper choice of the parameterization, a good description of
  binding energies and their differences is obtained, comparable to that for even
  nuclei. The description of low-energy excitation spectra of odd nuclei is of varying
  quality depending on the nucleus.
} 
\maketitle

\section{Introduction}

Nuclear density-functional-theory (DFT) in its most prominent versions,
Skyrme-Hartree-Fock (SHF), Gogny, and relativistic mean-field (RMF), are widely
used theoretical tools for the description of ground states and dynamics of
atomic nuclei, see the reviews
\cite{Bender_RMP_03_rew,Vretenar_PR_05_rew,Stone_PPNP_07_rew}. 
They have reached a high level of
quality in description of nuclear properties throughout the nuclear chart and,
what is important, do this in a self-consistent manner. The overwhelming
majority of applications deal with even-even nuclei which are far simpler to
handle than odd ones. Odd nuclei add crucial pieces of new information, e.g.
single-particle spectra, and thus are increasingly studied in recent years.
A fully self-consistent calculation for odd nuclei
has to account for all the terms following from the initial functional.  This
means, in particular, that contributions of both time-even and time-odd
densities from the initial functional are to be used
\cite{Bender_RMP_03_rew,Vretenar_PR_05_rew,Stone_PPNP_07_rew,Engel_NP_75,Dob_PRC_95}.
The time-odd densities (current $\boldsymbol j$, spin $\boldsymbol s$, and
vector kinetic-energy $\boldsymbol T$) and related time-odd mean fields (TOMF)
do not contribute to the ground state of non-rotating even-even nuclei but can
be essential for the ground states of odd and odd-odd nuclei, as well as in
nuclear dynamics including nuclear rotation
\cite{Dob_PRC_95,Afa_Ring_PRC_00,Zdu_PRC_05,Afa_PRC_08} and electric
\cite{nest_PRC_02,nest_PRC_06,nest_PRC_08,nest_IJMP_08}, and magnetic
\cite{nest_PRC_09_M1,nest_JPG_10_M1} giant resonances.  Moreover, TOMF embrace
the crucial terms to restore the local gauge invariance (Galilean invariance
in non-relativistic Skyrme and Gogny models) violated by velocity-dependent
time-even densities \cite{Engel_NP_75,Dob_PRC_95}.

In odd nuclei, TOMF contribute to the ground state mean fields
and, by breaking the time-reversal symmetry in the intrinsic frame, destroy the Kramer's
degeneracy of the single-particle states. TOMF might also influence
the single-particle spectra and such characteristics as binding
energies, odd-even staggering, and separation energies. Note that
binding energies and separation energies are important for
astrophysical applications \cite{Stone_PPNP_07_rew,Langanke_RPP_01},
the odd-even staggering determines the quality of the pairing
description
\cite{Dob_SkP,Rutz_NPA_98,Satula_99,Bender_EPJA_00,Duguet_PRC_01,Bertulani_PRC_09,Bertsch_PRC_09},
the separation energies are crucial for estimation of nuclear
stability in drip line \cite{Penion_PPN_06} and super-heavy regions
\cite{Ren_PRC_03}, and the low-energy single-particle spectra are now
in focus of exploration of exotic nuclei
\cite{Penion_PPN_06,Ahmad_PRC_05}.

The role of TOMF in odd nuclei has been already thoroughly inspected for
RMF models, see e.g. the recent study \cite{Afa_PRC_10_RMF} and references
therein. Most applications of SHF models neglect TOMF. There are a few early
publications \cite{Satula_99,Passler_NPA_76} taking care of TOMF, however,
involving only light nuclei and a limited number of Skyrme forces (SIII
\cite{SIII} and SLy4 \cite{SLy46}). The large interest on a proper description
of odd nuclei motivates a fresh look at the case. Thus very recently, a new
SHF study of rare-earth odd nuclei has been published \cite{Sch10a}, where the TOMF effect on
excitation spectra of odd nuclei was discussed in detail. Altogether the
previous studies have found rather weak influence of TOMF on properties of odd nuclei.
The maximal effect was obtained in light nuclei \cite{Satula_99,Afa_PRC_10_RMF}.
For binding energy, the TOMF were shown to give more binding within RMF
\cite{Afa_PRC_10_RMF} and effects of both sign within SHF \cite{Satula_99}.
The TOMF-induced changes in the single-particle spectra
were found very small \cite{Afa_PRC_10_RMF,Sch10a}. It was claimed
that RMF results much less depend on the force
parameterization than SHF ones \cite{Afa_PRC_10_RMF}.

The previous studies were limited by particular mass regions or specific
features of odd nuclei. At the same, it would be very instructive
to have a general view of the problem, covering most of the nuclear chart
and involving all the basic characteristics which might be affected
by TOMF. Just this aim is pursued in the present paper devoted to a
systematic exploration of TOMF effects in SHF calculations for proton and
neutron odd nuclei.
The study covers about 1282 odd nuclei with $16 \le Z \le 92$
(from sulfur to uranium) where experimental energy data are available.
Both spherical and axially deform nuclei are involved.
Three Skyrme parameterizations SkI3 \cite{ski3}, SLy6 \cite{SLy46},
and SV-bas \cite{svbas}
with essentially different effective masses ($m^*/m$=0.58,
0.69, and 0.9, respectively) are applied.
A variety of observables (binding energies, odd-even staggering of energies,
separation energies, and low lying single-particle spectra) are analyzed.
To separate the effects from time-odd spatial current $\textbf j$ and spin density
$\textbf s$, the time-odd terms in the Skyrme functional are sorted into
the minimal set for restoring Galilean invariance, and the sets with additional spin
and spin-gradient couplings.
The results are compared with previous studies in SHF
\cite{Satula_99,Sch10a} and RMF \cite{Afa_PRC_10_RMF}.

The paper is outlined as follows. In Sec. \ref{sec:model}, the
theoretical framework and calculation scheme are outlined. In particular,
the Skyrme functional with the time-odd terms and the pairing functional
are presented. Results of the calculations are discussed in
Sec. \ref{sec:result}. The summary is given in Sec. \ref{sec_concl}.
Some details concerning the functional parameters and single-particle wave
functions are done in Appendices A and B.

\section{Model and details of calculations}
\label{sec:model}
The starting point is the total energy $E$ of system
\cite{Bender_RMP_03_rew,Stone_PPNP_07_rew}
\begin{eqnarray}\label{tot_en}
   {E}  &=&  \int d \textbf{r}\Big\{ {\cal H}_{\rm kin}(\tau_q)
 +{\cal H}_{\rm Sk}(\rho_q,\tau_q,
   \textbf{s}_q,\textbf{j}_q,\textbf{J}_q,\textbf{T}_q)
\\
  && \qquad  + \ {\cal H}_{\rm pair}(\chi_q) + {\cal H}_{\rm C}(\rho_p)
    \Big\} -E_\mathrm{cm} \; ,
\nonumber
\end{eqnarray}
involving local time-even  (nucleon $\rho_q$, kinetic-energy  $\tau_q$,
spin-orbit  $\textbf{J}_q$) and time-odd (current $\textbf{j}_{ q}$, spin
$\textbf{s}_q$, spin kinetic-energy $\textbf{T}_q$) densities, as well as
the pairing density $\chi_q$. The label $q$ denotes protons and neutrons.

Total densities, like $\rho = \rho_p + \rho_n$, have no the index.
Further
\begin{eqnarray}
   {\cal H}_{\rm kin} &=&   \sum_{q}\frac{\hbar^2}{2m_q} \tau_q (\boldsymbol{r})\;,
 \label{Ekin}
\\
 {\cal H}_{\rm pair}
  &=&
  \frac{1}{4}F(\boldsymbol{r})\sum_q V_{{\rm pair},q}
  \chi^*_q (\boldsymbol{r}) \chi^{\mbox{}}_q (\boldsymbol{r})
\;,
\label{eq:pair_functional}
\\
  {\cal H}_{\rm C}
 &=&
  \frac{e^2}{2}
 \int  d{\boldsymbol r'} \rho_p(\boldsymbol{r})
              \frac{1}{|\boldsymbol{r}-\boldsymbol{r}'|} \rho_p(\boldsymbol{r}')
\nonumber\\
     &&
     -\frac{3}{4} e^2\left(\frac{3}{\pi}\right)^\frac{1}{3}
                          [ \rho_p(\boldsymbol{r})]^\frac{4}{3} \;,
\label{eq:Ecoul}
\\
  E_\mathrm{cm}
  &=&
  \frac{\langle\hat{\mathbf{P}}^2_\mathrm{cm}\rangle}{2mA}
   \;,
\label{eq:Ecm}
\end{eqnarray}
are kinetic-energy, pairing, Coulomb, and center-of-mass terms.
Note that the center-of-mass energy $E_\mathrm{cm}$ is not included
in the mean-field equations but evaluated a posteriori \cite{svbas}.
The pairing interaction may be density dependent
(surface pairing with $F=1- \rho/\rho_0$
and $\rho_0 = 0.20113$ fm$^{-3}$)
or not (volume pairing with $F=1$) \cite{Bender_EPJA_00}.

The key part of (\ref{tot_en}) is the Skyrme functional ${\cal H}_{\rm Sk}$
composed from the terms

\begin{eqnarray}
  \mathcal{H}_\mathrm{Sk}^\mathrm{(even)}
  &=&
  \frac{b_0}{2} \rho^2- \frac{b'_0}{2} \sum_q\rho_{q}^2
  \nonumber
  \\
 && +b_1 \rho \tau 
 - b'_1 \sum_q\rho_q \tau_q 
\nonumber
\\
 && -\frac{b_2}{2} \rho\Delta \rho
 + \frac{b'_2}{2} \sum_q \rho_q \Delta \rho_q
 \nonumber
 \\
  && +\frac{b_3}{3} \rho^{\alpha+2}
  - \frac{b'_3}{3} \rho^{\alpha} \sum_q \rho^2_q
\nonumber
\\
&& -b_4 \rho \nabla\textbf{J}
   -b'_4 \sum_q \rho_q \nabla\textbf{J}_q
\nonumber
\\
  &&
  -\tilde{b}_1\textbf{J}^2
  -\tilde{b}'_1   \sum_q\textbf{J}_q^2\;,
\label{eq:Sk-even}
\\
  \mathcal{H}_\mathrm{Sk}^\mathrm{(Gal)}
  &=&
  -b_1   \textbf{j}^2
  + b'_1   \sum_q \textbf{j}_q^2
\nonumber
\\
  &&
  -b_4(\nabla\!\times\!\textbf{j})\!\cdot\!s
  -b'_4 \sum_q
    (\nabla\!\times\!\textbf{j}_q)\!\cdot \!\textbf{s}_q
\\
  &&
  +\tilde{b}_1   \textbf{s} \!\cdot\!\textbf{T}
  + \tilde{b}'_1   \sum_q \textbf{s}_q\!\cdot\!\textbf{T}_q \;,
\nonumber
\label{eq:Sk-Gal}
\\
  \mathcal{H}_\mathrm{Sk}^\mathrm{(\textbf{s}^2)}
  &=&
  \frac{\tilde{b}_0}{2} \textbf{s}^2
  - \frac{\tilde{b}'_0}{2} \sum_q \textbf{s}_{q}^2
\nonumber
\\
 &&
   +\frac{\tilde{b}_3}{3} \rho^{\alpha} \textbf{s}^2
   - \frac{\tilde{b}'_3}{3} \rho^{\alpha} \sum_q \textbf{s}^2_q
   \;,
\label{eq:Sk-s}
\\
  \mathcal{H}_\mathrm{Sk}^\mathrm{(\textbf{s}\Delta\textbf{s})}
  &=&
  -\frac{\tilde{b}_2}{2} \textbf{s} \!\cdot\!
  \Delta \textbf{s} + \frac{\tilde{b}'_2}{2}
  \sum_q \textbf{s}_q \!\cdot\!\Delta \textbf{s}_q
\label{eq:skyrme_funct}
\end{eqnarray}
where $b_i$, $b'_i$, $\tilde{b}_i$, $\tilde{b}'_i$ are the force parameters.
Their connection with the standard Skyrme parameters $t_i$ and $x_i$
is specified in the Appendix \ref{app:params}.
The dominant Skyrme contribution $\mathcal{H}_\mathrm{Sk}^\mathrm{(even)}$ involves
only time-even densities $\rho_q$, $\tau_q$, and $\textbf{J}_q$.  It
embraces all standard terms relevant for the ground state properties and electric
excitations of even-even nuclei (for electric modes the current $\textbf{j}_q$
can be also important \cite{nest_PRC_02,nest_PRC_06,nest_IJMP_08}).
The isovector spin-orbit interaction is
usually linked to the isoscalar one by $b'_4=b_4$ and the tensor spin-orbit
terms $\propto\tilde{b}_1,\tilde{b}'_1$ are often omitted
\cite{Bender_RMP_03_rew,Stone_PPNP_07_rew}.
For the sake of simplicity we
also omit them and the present study and so use
$\tilde{b}_1=0$ and $\tilde{b}'_1=0$ throughout.

The next Skyrme contributions,
$ \mathcal{H}_\mathrm{Sk}^\mathrm{(Gal)},
\mathcal{H}_\mathrm{Sk}^\mathrm{(\textbf{s}^2)},
\mathcal{H}_\mathrm{Sk}^\mathrm{(\textbf{s}\Delta\textbf{s})}$,
add gradually terms with the time-odd densities
$\textbf{j}_{ q}$, $\textbf{s}_q$, and $\textbf{T}_q$ and create
corresponding TOMF in the mean-field equations.  These are the terms of
our particular interest.  They do not contribute to ground state properties
of even-even nuclei but become relevant in odd nuclei and nuclear dynamics
(rotation, electric and magnetic giant resonances).
  In the present study we consider three options for the TOMF impact:
\begin{eqnarray}
  &\mathcal{H}&_\mathrm{Sk}^\mathrm{(even)}
  ,
\label{eq:op1}
\\
  &\mathcal{H}&_\mathrm{Sk}^\mathrm{(min)}
  =
  \mathcal{H}_\mathrm{Sk}^\mathrm{(even)}
  +
  \mathcal{H}_\mathrm{Sk}^\mathrm{(Gal)}
  \;,
\label{eq:op2}
\\
  &\mathcal{H}&_\mathrm{Sk}^\mathrm{(min+\mathbf{s}^2)}
  =
  \mathcal{H}_\mathrm{Sk}^\mathrm{(even)}
  +
  \mathcal{H}_\mathrm{Sk}^\mathrm{(Gal)}
  +
  \mathcal{H}_\mathrm{Sk}^\mathrm{(\mathbf{s}^2)}
  \;.
\label{eq:op3}
\end{eqnarray}
The first option deals only with $\mathcal{H}_\mathrm{Sk}^\mathrm{(even)}$,
i.e. involves only time-even densities.
This violates Galilean invariance in nuclear dynamics
\cite{Engel_NP_75}. Adding the term $\mathcal{H}_\mathrm{Sk}^\mathrm{(Gal)}$
restores this invariance. This yields
$\mathcal{H}_\mathrm{Sk}^\mathrm{(min)}$, the minimal set to achieve a
consistent dynamical model. Derivation of the functional from a Skyrme force
\cite{Skyrme,Vauterin} also yields the spin terms
$\mathcal{H}_\mathrm{Sk}^\mathrm{(\textbf{s}^2)}$ and
$\mathcal{H}_\mathrm{Sk}^\mathrm{(\textbf{s}\Delta\textbf{s})}$
with the parameters
$\tilde{b}_i$ and $\tilde{b}'_i$ uniquely related to the $b_i$, $b'_i$, see
Appendix \ref{app:params}.  Following our experience, the term
$\mathcal{H}_\mathrm{Sk}^\mathrm{(\mathbf{s}\Delta\mathbf{s})}$ leads to
unstable ground states for many medium and heavy nuclei. This is a
common problem in most Skyrme parameterizations \cite{Les06a,Kor10a}.
Thus we include to the option $\mathcal{H}_\mathrm{Sk}^\mathrm{(min+\mathbf{s}^2)}$
only the simple spin term $\mathcal{H}_\mathrm{Sk}^\mathrm{(\textbf{s}^2)}$.
The comparison of results for $\mathcal{H}_\mathrm{Sk}^\mathrm{(even)}$ with
those from $\mathcal{H}_\mathrm{Sk}^\mathrm{(min)}$ and
$\mathcal{H}_\mathrm{Sk}^\mathrm{(min+\mathbf{s}^2)}$ should provide information
on the influence of TOMF. The corresponding total energies are denoted as
$E_\mathrm{even}$, $E_\mathrm{min}$, and $E_\mathrm{min+\mathbf{s}^2}$.

The calculations are performed in cylindrical coordinate space grid with the
mesh size 1 fm. The axial equilibrium deformations are determined by
minimization of the total energy \cite{Bender_RMP_03_rew}. The calculations are
restarted from different initial deformations to separate isomeric and
absolute minima of the energy.

To explore the possible variance of the results, the calculations are done
for three different Skyrme parameterizations, SkI3 \cite{ski3}, SLy6 \cite{SLy46},
and SV-bas \cite{svbas} with the effective masses $m^*/m$=0.58, 0.69, and 0.9,
respectively. All the parameterizations provide a reasonable description of stable
even-even nuclei and giant resonances, though perform differently with respect to more
detailed observables. The force SkI3 is known to reproduce peculiarities of Pb
isotopic chain. 
SLy6 was developed with special emphasis on neutron rich nuclei and neutron matter.
The recent parameterization SV-bas \cite{svbas} is
fitted for a large data set covering long isotopic and isotonic chains and so may be 
most suited for systematic explorations.
Note that  SkI3 and SV-bas allow an isovector
spin-orbit force (decoupled parameter $b'_4$) while SLy6 fixes that
to the isoscalar term by setting $b'_4=b_4$. Both these differences can affect the
single-particle shell structure and so the properties of odd nuclei.

The pairing functional (\ref{eq:pair_functional}) uses the volume pairing
for SkI3 and SLy6 and surface pairing for SV-bas. The pairing strengths
$V_{{\rm pair},q}$ are adjusted following the recipe \cite{svbas}.
The pairing is augmented by the stabilization procedure from \cite{Erl08a}. This prevents
the breakdown of the pairing near the closed shells which to some extent
simulates the similarly smoothing in particle-number projection and
so avoids unphysical kinks in properties along collective deformation paths.

The pairing mean-field equations are derived from the
given pairing functional and solved in the BCS approximation.
The pairing density in (\ref{eq:pair_functional}) is
\begin{equation}\label{eq:pair_dens}
  \chi_q(\textbf{r})=-\sum_{k\in q} u_k v_k f_k^q \sum_{\sigma = \pm} \sigma
\phi_{\bar{k}}^{(-\sigma)}(\textbf{r})\phi_{k}^{(\sigma)}(\textbf{r})
\end{equation}
where the sum runs over the single-particle states $k$,
the state $\bar{k}$ is time-conjugate  to $k$,
$v_k$ and $u_k$ are Bogoliubov amplitudes,
$\phi_{k}^{(\sigma)}(\textbf{r})$ is the spin $\sigma$-component of the
single-particle wave function in cylindrical coordinates (see appendix
\ref{app:spwf}).

The weight in (\ref{eq:pair_dens})
\begin{equation}\label{cut-off_factor}
  f_k^q=\frac{1}
      {1+\text{exp}
      \left\{[0.5(\epsilon_k+\epsilon_{\bar{k}})- \lambda_q - \Delta E_q]/\mu_q\right\}}
\end{equation}
is an energy-dependent cut-off factor limiting the pairing space to the
vicinity of the Fermi energy. The cut-off is necessary to suppress
unphysically large contributions from high-lying states, caused by the
zero-range form of the pairing interaction \cite{Bender_RMP_03_rew}.
We use here the recipe \cite{Bender_EPJA_00} with the cut-off parameters
amounting in average to the typical values $\Delta E_q$=5 MeV and
$\mu_q$=0.5 MeV \cite{Kri90a}.
The value $\epsilon_k$
is the single-particle energy of state $k$ and $\lambda_q$ is the chemical
potential.

In odd nuclei, the odd nucleon misses a partner to join the pairing
scheme and thus blocks the occupied single-particle state.
The blocking results in the one-quasiparticle state
\cite{Ring_Schuck_book,Solov_book}
\begin{equation}
\hat \alpha^+_{k_1}|\text{BCS}\rangle=
\hat a^+_{k_1}\prod_{k>0, k \ne k_1} (u_k + v_k \hat a^+_k \hat a^+_{\bar k})|0\rangle \; ,
\label{eq:block}
\end{equation}
where $\hat \alpha^+_{k}$ and $\hat a^+_{k}$ are quasiparticle and single-particle
creation operators;
$|\text{BCS}\rangle$ and $|0\rangle$ are BCS and HF vacuum states.
The blocking effect is known to be of a crucial importance for odd nuclei
\cite{Ring_Schuck_book,Solov_book,Gareev_NP_71}.

We compute the blocked states for a
variety of single-particle states $k$ near the Fermi energy. The state
with lowest energy is treated as the ground state while others
form the low-energy excitation spectrum of the odd nucleus.

Finally note that time-conjugate states are easily
obtained in even-even nuclei as $\phi_{\overline k}=\hat T\phi_k$
by using the time-reversal operator $\hat T$.
However, this is not trivial for odd nuclei where the time-reversal
invariance is broken. In this case,
$\phi_{\overline k}$ is found as the state having maximum overlap
with $\hat T\phi_k$.

\section{Results and discussion}
\label{sec:result}

In this section we present our results for the influence of TOMF on
binding energies, odd-even staggering, separation energies, and
low-energy spectra of odd nuclei.  Three options (\ref{eq:op1})-(\ref{eq:op3})
with the corresponding total ground state energies, $E_\mathrm{even}$,
$E_\mathrm{min}$, and $E_\mathrm{min+\mathbf{s}^2}$, are compared.
The same index convention is applied to other observables.

In this study we neglect the coupling with vibrational states of the
even-even core \cite{Solov_book,Gareev_NP_71} and nuclear rotation
\cite{Alikov_ZPA_88,nest_odd_93}.  The loss of the corresponding
collective correlations can result in underestimation of the total
binding energies by 1-2 MeV in soft or deformed nuclei
\cite{Klu08c,Gor01a}.  For the variables represented through the
differences of the energies (pairing gaps and separation energies) the
effect should be much smaller.  For single-particle energies the role
of the vibrational and rotational coupling can be dramatic
\cite{Alikov_ZPA_88}. In general it is reduced to much more compressed
low-energy spectrum of odd nuclei \cite{Solov_book,Gareev_NP_71}. So
we may expect from very beginning for more dilute SHF spectra as
compared to the experimental data.

\subsection{Binding energies}

\begin{figure}[t]
\includegraphics[width=\linewidth]{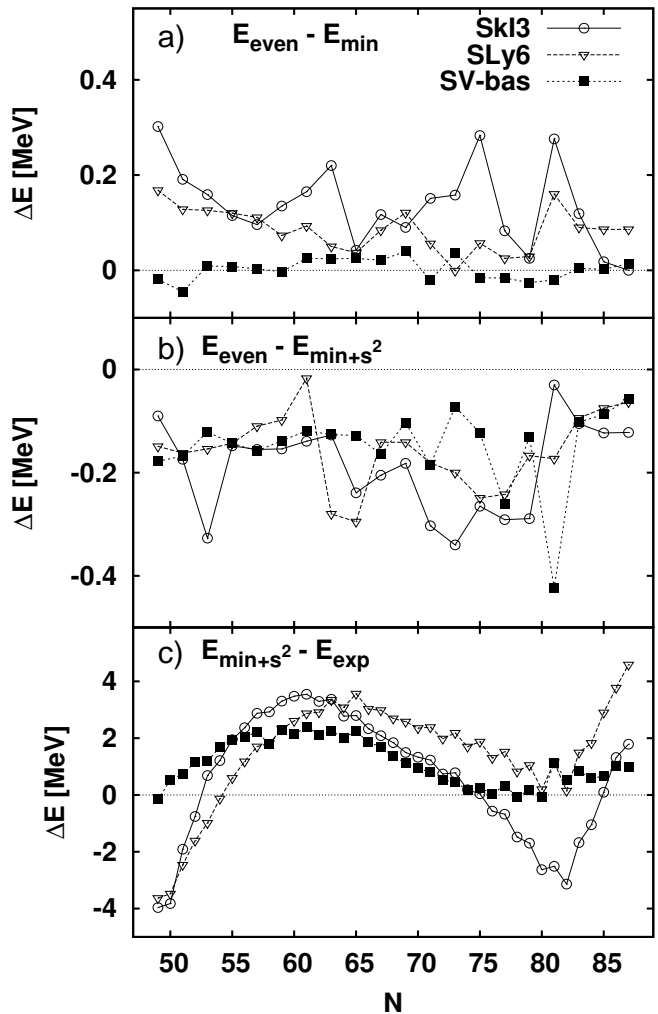}
\caption{
\label{fig1}
The energy differences
$E_\mathrm{even}-E_\mathrm{min}$ (a),
$E_\mathrm{even}-E_\mathrm{min+\textbf{s}^2}$ (b), and
$E_\mathrm{min+\textbf{s}^2}-E_{\mathrm{exp}}$ (c),
in odd Sn isotopes for the parameterizations SkI3, SLy6, and SV-bas.}
\end{figure}

\begin{figure}[t]
\includegraphics[width=\linewidth]{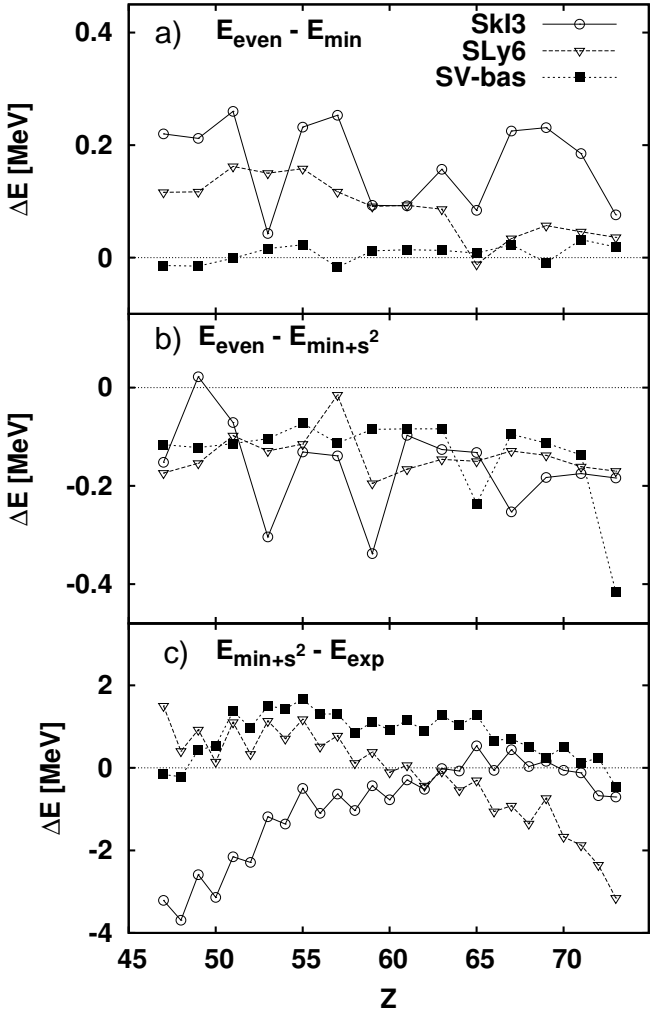}
\caption{
\label{fig2}
The same as Fig. \ref{fig1}, but for N=82 isotones.}
\end{figure}

\begin{figure}
\includegraphics[width=\linewidth]{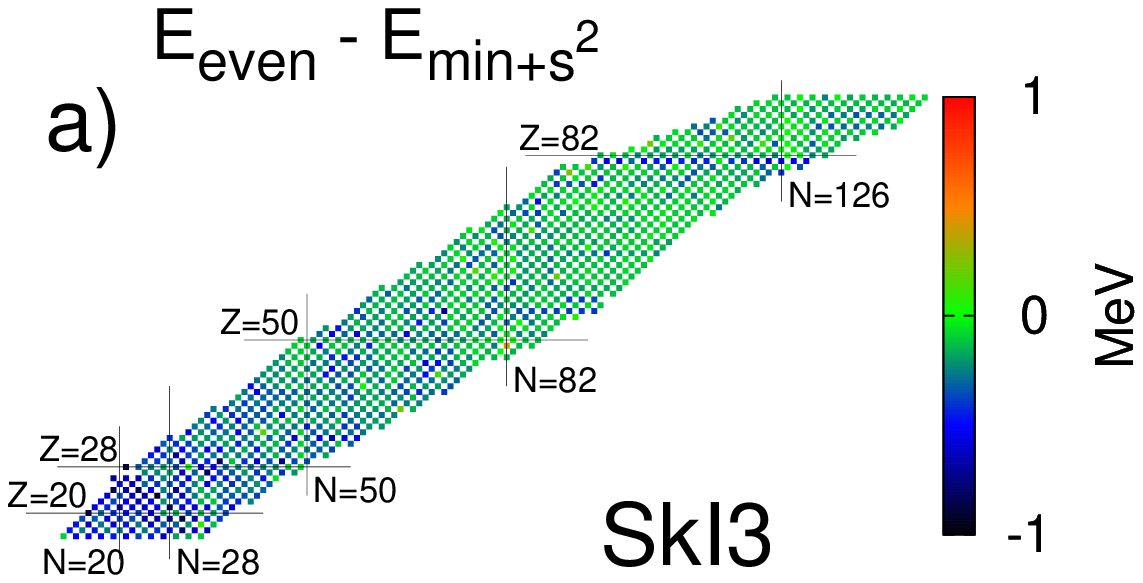}
\includegraphics[width=\linewidth]{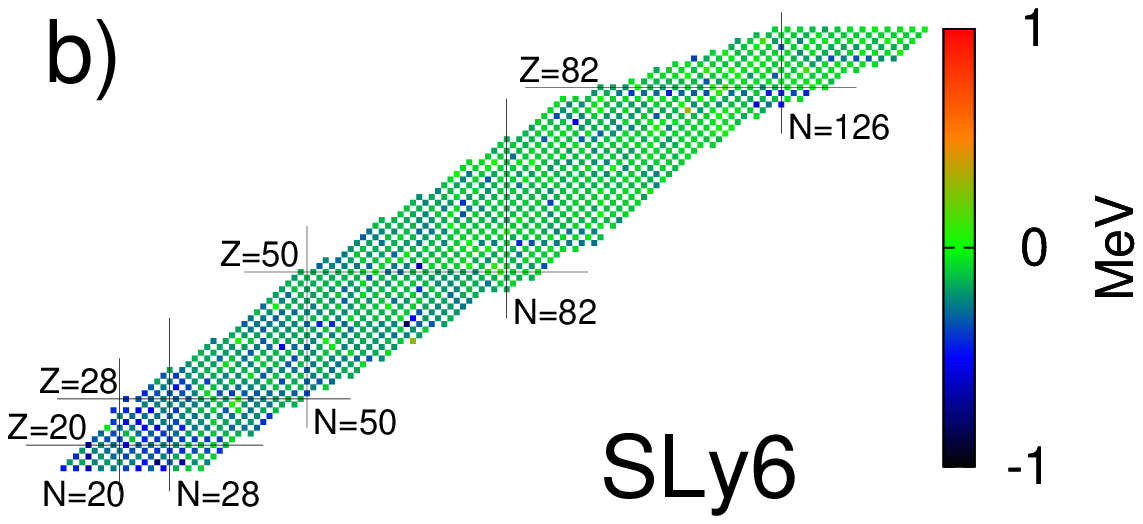}
\includegraphics[width=\linewidth]{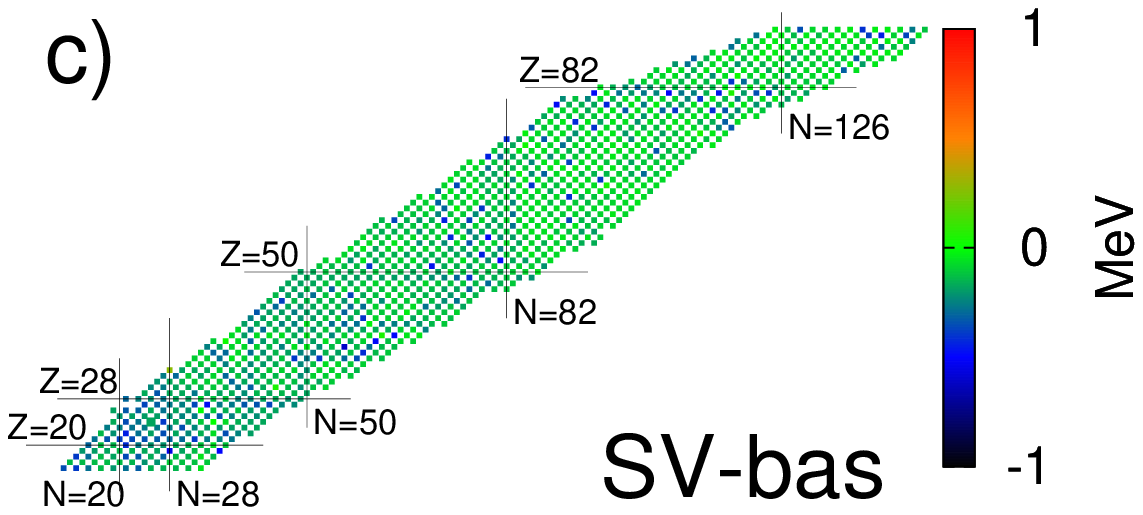}
\caption{
\label{fig3}
The binding energy difference
$E_\mathrm{even}-E_\mathrm{min+\textbf{s}^2}$ for the Skyrme
parameterizations SkI3, SLy6 and SV-bas, drawn in the $N$-$Z$
plane. All available neutron and proton
odd nuclei with the charge $16 \le Z \le 92$ are included.}
\end{figure}

\begin{figure}[t]
\includegraphics[width=\linewidth]{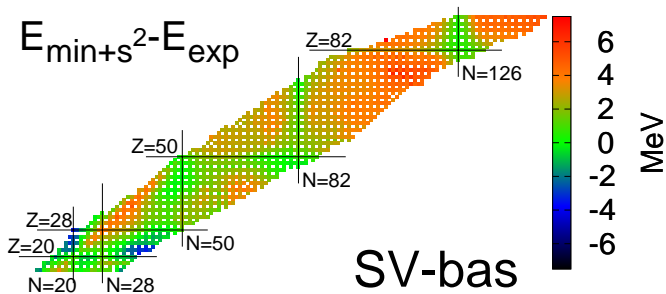}
\caption{
\label{fig4}
The difference between SHF and experimental values \protect\cite{exp_BE} of
the binding energy, $E_\mathrm{min+\textbf{s}^2}-E_\mathrm{exp}$, calculated
for the force SkI3 and plotted in the $N$-$Z$ plane. All even and odd nuclei
with $16 \le Z \le 92$ are included.}
\end{figure}

Results for the binding energies (BE) are given in
Figs. \ref{fig1}-\ref{fig4} and Table 1. The experimental data are taken
from \cite{exp_BE}.
To amplify the effects,
the energy differences $\Delta E$ are considered:
$E_\mathrm{even}-E_\mathrm{min}$ to check the effect of restoration
of Galilean invariance,
$E_\mathrm{even}-E_\mathrm{min+\mathbf{s}^2}$ to demonstrate the total
TOMF effect when spin terms $\mathbf{s}^2$ are also added, and
$E_\mathrm{min+\mathbf{s}^2}-E_\mathrm{exp}$ to estimate the
deviation from the experimental data.

Fig. \ref{fig1} shows the BE differences for odd Sn isotopes. Following
panel a), the restoration of Galilean invariance enhances the binding.
The effect is small throughout and practically negligible for SV-bas. This
is not surprising as SV-bas with its effective mass $m^*/m=0.9$ has smaller
coefficients $b_1$ and $b'_1$ than SLy6 and SkI3 and hence weaker current
contribution $\sim \mathbf{j}^2$. The additional inclusion of
spin terms (panel b) counterweights the extra binding and leads to its modest
reduction by 0.1-0.3 MeV. Thus the effects of TOMF on
the BE are generally small.

Panel c) of Fig. 1 shows the difference
$E_\mathrm{min+\textbf{s}^2}-E_\mathrm{exp}$ between theory and experiment
in both even and odd Sn isotopes.  This difference is much larger than the
previous ones and shows remarkable isotopic
trends.  The typical patterns for SkI3 and SLy6 show a tendency to over-binding
at the shell closures N=50 and 82 and under-binding at mid shell.  The most recent
parameterization SV-bas produces the smoothest trends and smallest
deviations extending to the limits of known isotopes. This is due including
the long isotopic and isotonic chains into the SV-bas fit \cite{svbas}.
Even here there remains a region of under-binding by $\sim$ 2 MeV near
N=55-65. But these nuclei are soft and, as was mentioned above, should additionally
acquire just this amount of the collective correlation energy \cite{svbas,Klu08c}.
Anyway, we see large deviation from the experiment and strong dependence of the results
on the Skyrme parameterization. Both these factors are an order of magnitude stronger
than the TOMF effects shown in panels a) and b).

Fig. \ref{fig2} shows the BE differences for the chain of isotones
with the neutron number N=82. Most of the patterns are much the same
as for the isotopic chain in Fig. \ref{fig1} and so the above remarks
can be applied for Fig. \ref{fig2} as well. The only exception seems
to be a different behavior of $E_\mathrm{min+\textbf{s}^2}-E_\mathrm{exp}$
for SkI3 and SLy6 in panels c) of Figs. 1 and 2, where we see
similar SkI3 and SLy6 trends for isotopes and opposite for isotones.
Perhaps, this is related with different fit strategies of these two
parameterizations.

In the following we discuss nuclei all over the chart of isotopes, 
even-even nuclei and proton-odd or neutron-odd ones called henceforth 
``odd''. Doubly odd nuclei are not considered.

Fig. \ref{fig3} shows the
difference $E_\mathrm{even}-E_\mathrm{min+\textbf{s}^2}$ all over the
nuclear chart with $16 \le Z \le 92$.
The effect of TOMF is weak everywhere, never exceeding 1 MeV and usually remaining
even much smaller.
In general the TOMF lead to less binding.  The effect is
stronger in light nuclei and decreases with the nuclear size, which is
in accordance with RMF  findings  \cite{Afa_PRC_10_RMF}.
There is some dependence on the Skyrme parameterization concerning the
increase of the TOMF effect for small nuclei, which is more pronounced
for SkI3 and less for SV-bas.  Note that, in contrast to the previous
Skyrme study \cite{Satula_99} for light nuclei, neither the RMF
\cite{Afa_PRC_10_RMF} nor our calculations find an enhancement of the
TOMF effect at N=Z. Perhaps in our case this is caused by omitting the
term $\textbf{s} \cdot \Delta \textbf{s}$ which was taken into account in
\cite{Satula_99} and gave there a strong contribution opposite to one
from $\textbf{s}^2$-terms.

Fig. \ref{fig4} shows the systematics of the deviation of calculated BE
from the experimental data \cite{exp_BE} in even and odd nuclei altogether. As
all three parameterization show qualitatively the same pattern, we
show only the results from SV-bas.  As might be expected, the best
agreement is found along the isotopic and isotonic chains of semi-magic
nuclei. These are the nuclei with least correlation effects \cite{Klu08c}.

The present calculations omit these effects and generally yield under-binding
for deformed nuclei.
Subtracting proper correlation energies cures the
problem for light and medium nuclei. But there remains a systematic
trend to under-binding of deformed heavy and super-heavy elements
which is a well known problem of all Skyrme parameterizations
\cite{svbas,Erl10a}. The same trends exist for even-even
nuclei and for other Skyrme parameterizations. Actually they are features
of the present SHF functionals as such. Most important for our purposes is
that odd nuclei do not change the picture at all. They would merge
smoothly into the data from even nuclei without showing any special
even-odd effect. Proper tuning of BE in even nuclei thus
automatically provides good results for BE in odd nuclei as well.

Additional information on the global quality of BE description
is done in Table \ref{tab:be_chi^2} in terms of r.m.s. deviation
from experimental data \cite{exp_BE}.
The numbers have to
be taken with care because they are computed from raw mean-field
energies without rotational projection and vibrational correlation
energy.

Table \ref{tab:be_chi^2} shows that restoration of the Galilean invariance
(step from $E_\mathrm{even}$ to $E_\mathrm{min}$) improves the description
only a bit. But adding the $\textbf{s}^2$-terms, i.e. stepping to
$E_\mathrm{min+\textbf{s}^2}$, again enhances the deviations. Note
that even and odd nuclei are described about with a similar
accuracy. SV-bas yields to a better description than SkI3 and SLy6
and, again, the difference between the Skyrme parameterizations is
much larger than the TOMF effect.

\begin{table}[t]
\caption{R.m.s. deviations (in MeV) between SHF and experimental data
\cite{exp_BE}  for the binding energies in odd, even,
and all (odd + even) nuclei from Figs. \ref{fig3} and \ref{fig4}.}
\label{tab:be_chi^2}
  \begin{tabular}{clccc}
   Parameterization & Option & \multicolumn{3}{c}{Nuclei}\\
   \hline
        &                 & even  & odd  & all \\
   \hline
        & $\text E_\mathrm{even}$&3.51&3.57&3.55\\
   SkI3 & $\text E_\mathrm{min}$&3.51&3.50&3.50\\
        & $\text E_\mathrm{min+\textbf{s}^2}$&3.51&3.73&3.66\\
  \hline
        & $\text E_\mathrm{even}$&4.18&4.01&4.07\\
   SLy6 & $\text E_\mathrm{min}$&4.18&3.97&4.04\\
        & $\text E_\mathrm{min+\textbf{s}^2}$&4.18&4.14&4.15\\
  \hline
          & $\text E_\mathrm{even}$&2.81&2.72&2.75\\
   SV-bas & $\text E_\mathrm{min}$&2.81&2.71&2.74\\
          & $\text E_\mathrm{min+\textbf{s}^2}$&2.81&2.84&2.83
  \end{tabular}
\end{table}

Finally it is worth to compare our results with the RMF predictions,
in particular with the recent study of \cite{Afa_PRC_10_RMF} covering
light nuclei with $ 18 \le Z \le 27$ as well as Ce and Cf isotopes.
Like in our case, they predict a weakening of the time-odd effects
with increasing nuclear size. At the same time, the RMF calculations
in light nuclei demonstrate a profound TOMF effect near the proton
drip-line and generally more binding induced by the odd-time
fields. These findings are not confirmed by our calculations, in
particular, the Skyrme calculation usually reduce binding when
including spin terms. However, we should not overestimate the
significance of this discrepancy.  In both RMF and Skyrme
calculations the TOMF contribution to the binding energies consists of
a terms of a different sign and actually the final result is determined
by a subtle balance of these terms. Using another parameterization or
adding still neglected contributions (like $\textbf{s} \cdot \Delta
\textbf{s}$ in the SHF) may change the balance
and so  magnitude and sign of a week TOMF effect.

In agreement to the previous SHF study \cite{Satula_99}, the present
calculations demonstrate a strong dependence of the results on the
parameterization.  This is to a large extent a consequence of using
essentially different parameterizations SkI3, SLy6, and SV-bas (which possess
much different effective masses and vary also in other features).  The RMF
results \cite{Afa_PRC_10_RMF}, on the other hand, are claimed to be almost
independent of the parameterization. But it is ought to mention that the RMF
parameterizations in this study are rather similar (all from group A in the
classification of \cite{Afa_PRC_10_RMF}). More differences may appear when
utilizing other RMF parameterizations which have higher
effective masses.

\subsection{Odd-even staggering (OES)}

During last years the odd-even staggering (OES) was a subject of
intense investigations within various models involving RMF
\cite{Rutz_NPA_98,Afa_PRC_10_RMF}, SHF
\cite{Dob_SkP,Satula_99,Bender_EPJA_00,Duguet_PRC_01,Bertulani_PRC_09,Bertsch_PRC_09},
and Gogny \cite{Dob_Gogny_96,Dob01d} approaches. Being mainly attributed to
the pairing, the OES is widely used as a measure of the pairing
gap. However, the OES can be also affected by other physical
mechanisms (e.g. nuclear deformation or TOMF)
which significantly
complicates the problem and calls for discrimination of the various
influences.  
For SHF, the TOMF effect on OES was so far explored either
indirectly \cite{Duguet_PRC_01} or only for light nuclei
\cite{Satula_99}. To the best of our knowledge, the present study is
the first direct and systematic investigation of the TOMF influence on
OES. The experimental values of OES used in the present exploration
are obtained by using evaluations \cite{exp_BE}.

\begin{figure}[t]
\includegraphics[width=\linewidth]{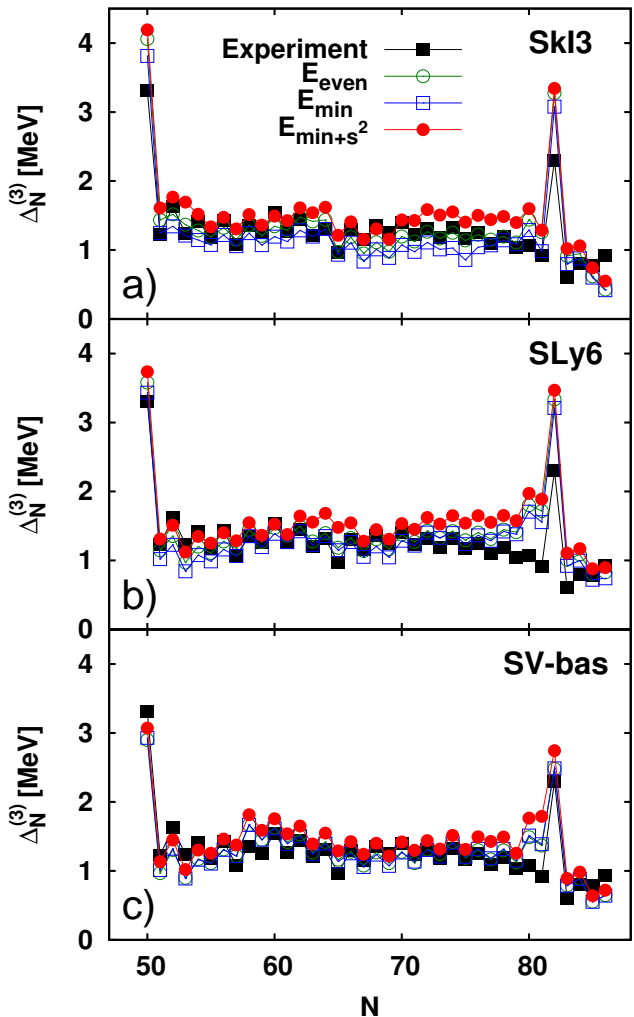}
\caption{
\label{fig5}
Neutron gaps $\Delta^{(3)}_{N}$ in Sn isotopes, calculated
for the parameterizations SkI3 (panel a), SLy6 (panel b),
and SV-bas (panel c) for the options (\ref{eq:op1})-(\ref{eq:op3})
as indicated and compared with the experiment \protect\cite{exp_BE}.}
\end{figure}
\begin{figure}[t]
\includegraphics[width=\linewidth]{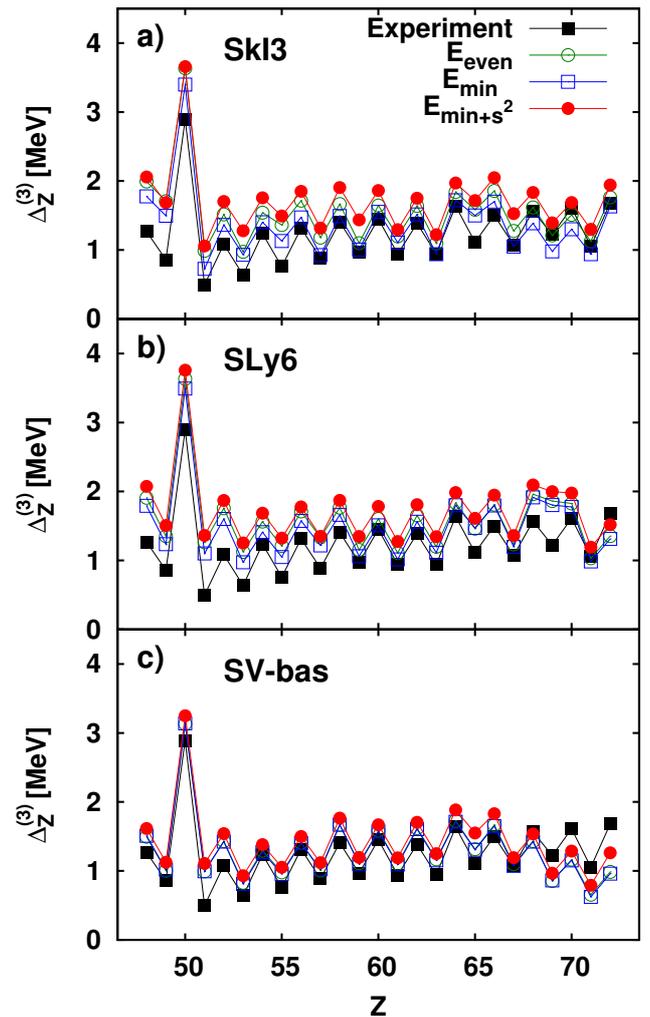}
\caption{
\label{fig6}
The same as in Fig. \ref{fig5} for the proton
gaps $\Delta^{(3)}_Z (Z,N)$ in N=82 isotones.}
\end{figure}

There exist several options for characterizing the OES \cite{Ring_Schuck_book}.
We adopt the three-point formulas
\begin{eqnarray}
  \Delta^{(3)}_N (Z,N) &=& \frac{(-1)^N}{2}[E(Z,N\!-\!1)-2E(Z,N)
\nonumber
\\
                      &&\qquad\quad +\ E(Z,N\!+\!1)]
                  \;,
\label{D3N}
\\
 \nonumber
 \Delta^{(3)}_Z (Z,N) &=& \frac{(-1)^Z}{2}[E(Z\!-\!1,N)-2E(Z,N)
 \\
                      &&\qquad\quad +\ E(Z\!+\!1,N)].
\label{D3Z}
\end{eqnarray}
It ought to be mentioned that background contributions act differently
for centering $\Delta^{(3)}_{Z,N}$ at even or odd nuclei. When centering  
at odd-N or odd-Z nucleus, respectively.
The three-point formulae
have the advantage of canceling the smooth background of mean field
contributions.  Instead, when being centered on even nuclei,
(\ref{D3N})-(\ref{D3Z}) are likely to contain also significant shell
and deformation effects \cite{Bender_EPJA_00,Dob01d}.

\begin{figure}[t]
\includegraphics[width=\linewidth]{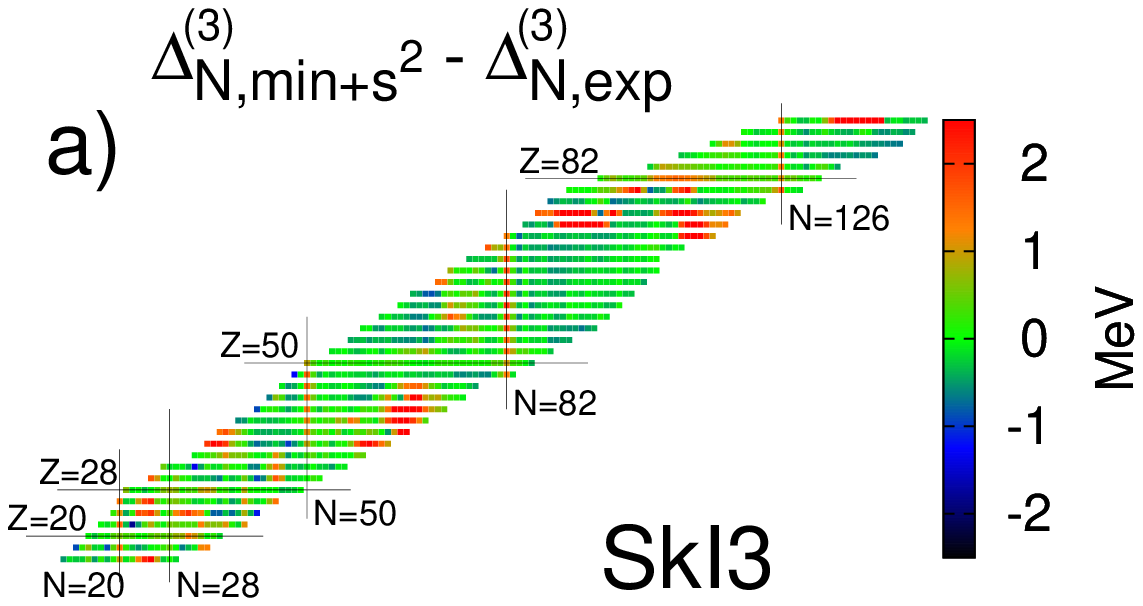}
\includegraphics[width=\linewidth]{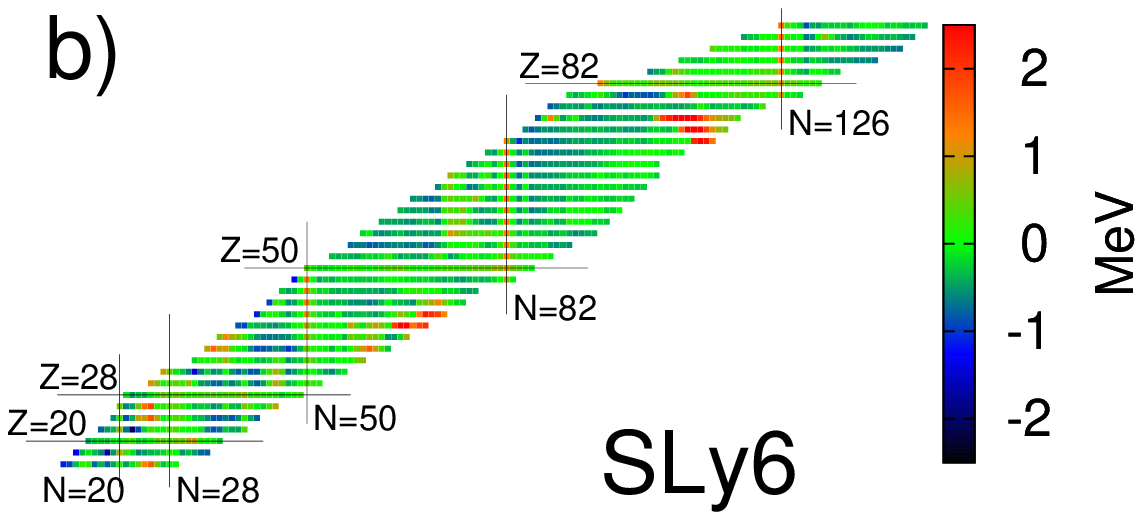}
\includegraphics[width=\linewidth]{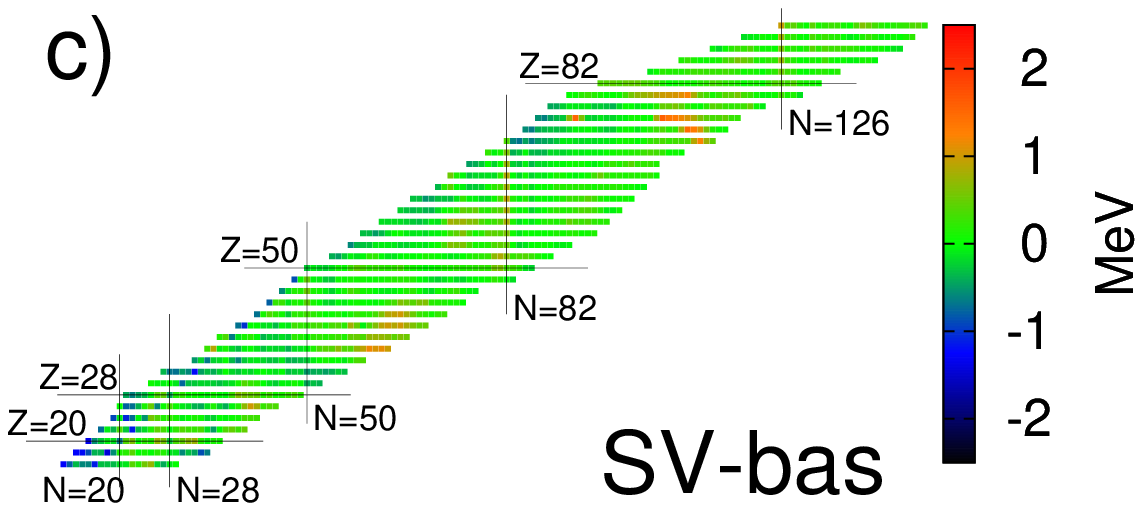}
\caption{
\label{fig7}
The difference between the theoretical and experimental
values \protect\cite{exp_BE} of the neutron pairing gaps,
$\Delta^{(3)}_{N, min+\textbf{s}^2} - \Delta^{(3)}_{N, exp}$,
for all available nuclei in the window  $16 \le Z \le 92$
and for the Skyrme parameterizations SkI3, SLy6, and SV-bas.}
\end{figure}
\begin{figure}[t]
\includegraphics[width=\linewidth]{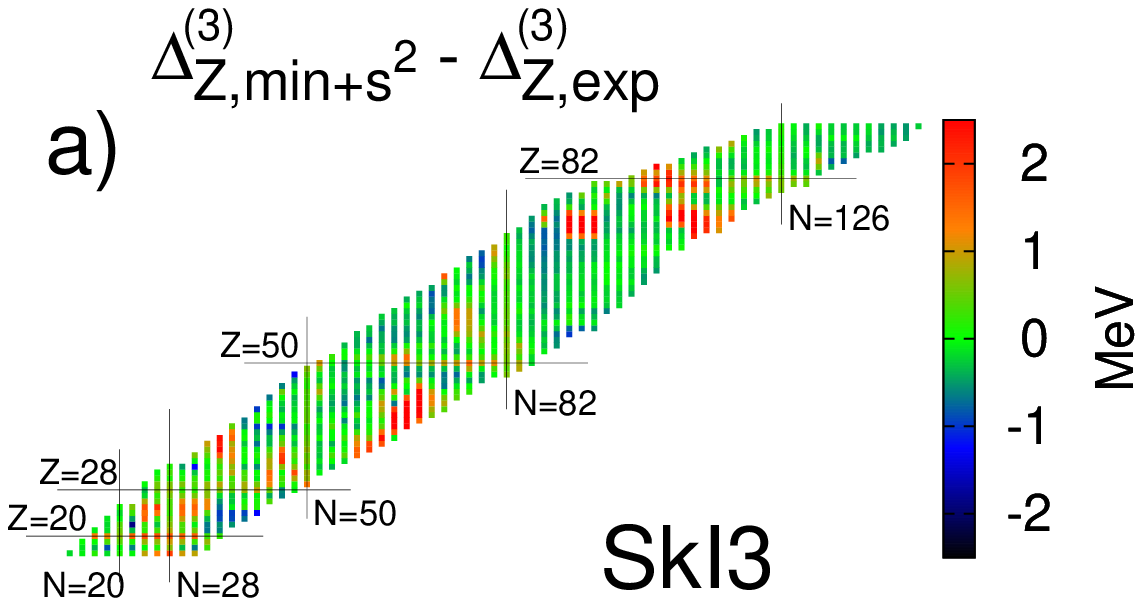}
\includegraphics[width=\linewidth]{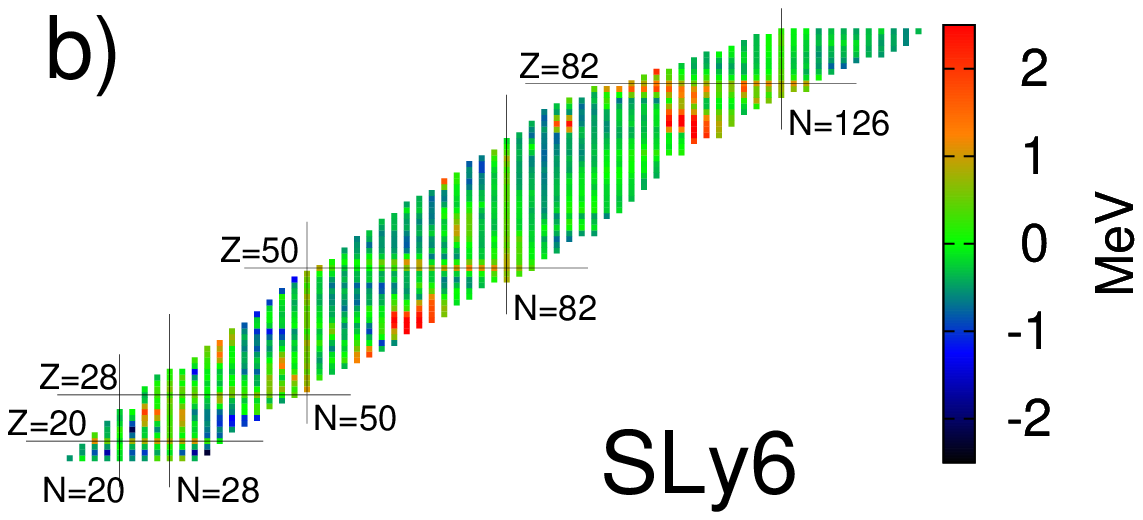}
\includegraphics[width=\linewidth]{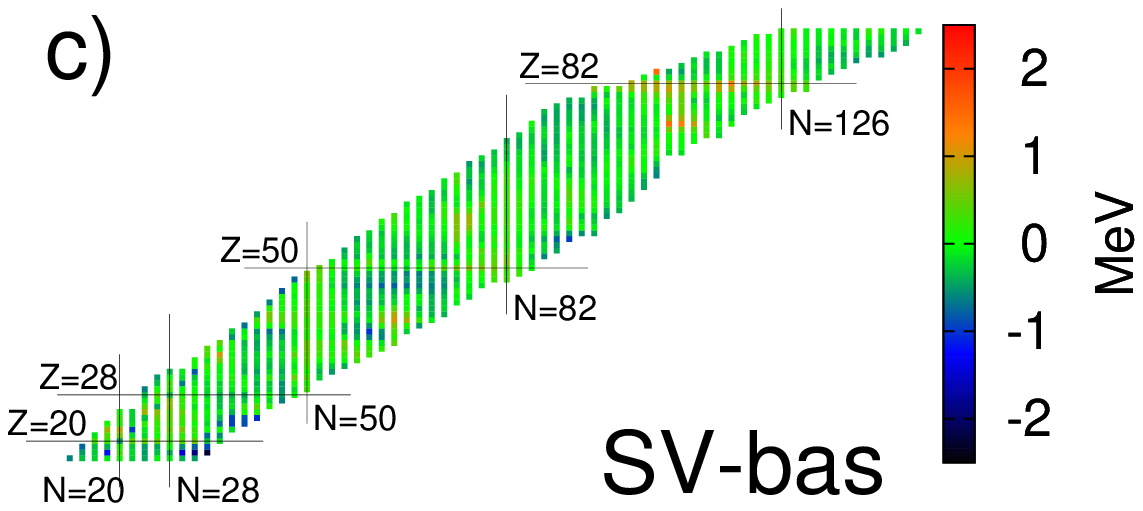}
\caption{
\label{fig8}
The same as in Fig. \ref{fig7} for the proton pairing gaps
$\Delta^{(3)}_Z (Z,N)$.}
\end{figure}

Fig. \ref{fig5} shows the neutron gaps along the chain of Sn
isotopes.
All three parameterizations, SkI3, SLy6, and SV-bas, quite
accurately describe the OES in sizes and trends.  The best agreement
is seen in the mid shell regions where pairing dominates in forming
the gaps and fit of the pairing strength is accomplished \cite{svbas}.
Larger deviations and a strong dependence on the parameterization are
found at and around the
shell closures (N=50 and 82). That is a point where shell effects
dominate and the shell structure varies substantially amongst the
parameterizations (e.g. due to the much different effective mass).
At a more detailed level, one can see that, for SkI3 and SLy6,
E$_\mathrm{min}$ slightly underestimates $\Delta^{(3)}_N$ while
E$_\mathrm{min+\textbf{s}^2}$ gives the opposite effect.  For SV-bas,
both E$_\mathrm{min}$ and E$_\mathrm{min+\textbf{s}^2}$ results are close to
experiment. The description is not so perfect for the N=82 isotones
shown in Fig. \ref{fig6} where all three SHF parameterizations show
somewhat larger deviations from the experimental data, especially
E$_\mathrm{min+\textbf{s}^2}$ for SkI3 and SLy6.  But even these deviations
are yet quite moderate and one may speak on a generally good description
of OES.  The trends like  a general decrease of $\Delta^{(3)}_N$
with N and increase of $\Delta^{(3)}_Z$ with Z are reproduced equally
well as in the study \cite{Bertulani_PRC_09} with the special isospin
pairing interaction.

Figs. \ref{fig7} and \ref{fig8} show the systematics of the deviations
between calculated $\Delta^{(3)}_N$ (Fig. \ref{fig7}) or
$\Delta^{(3)}_Z$ (Fig. \ref{fig8}) and the experimental data.
The majority of nuclei stays in the ``green'' window having perfect
agreement with the data. We see strings of larger deviations along
some semi-magic chains
and regions of well deformed and transitional nuclei.
These nuclei are known to have large collective correlations \cite{Klu08c}
and are likely to need them for improvement of the gap description.
The size of the deviations differs very much amongst the three
parameterizations and SV-bas again gives the best results.
This is most probably due to the large effective mass
($m^*/m=0.9$) which seems to produce a more correct level density.

The r.m.s. deviations for OES are summarized in Table
\ref{tab:OES_chi^2}.  Comparison with Table \ref{tab:be_chi^2} shows
that the average deviation for OES is much smaller than for the binding
energies, which reflects the fact that OES are less sensitive to
nuclear bulk properties and related errors. This is one more example
for the fact that energy differences are often more reliably described
than a total energy as such \cite{svbas}.
Like for the binding energies in Table \ref{tab:be_chi^2}, the TOMF effect on OES
is smaller than variations between the different
parameterizations.
\begin{table}
\caption{R.m.s. deviations (in MeV) between SHF and experimental data
\cite{exp_BE} for
$\Delta^{(3)}_{N,Z}$ from Fig. \ref{fig7} and
$\Delta^{(3)}_Z$ from Fig. \ref{fig8}.
}
\label{tab:OES_chi^2}
 \begin{center}
  \begin{tabular}{lccc}
   $\Delta^{(3)}_{N}$& SkI3 & SLy6  & SV-bas\\  \hline
   $\text E_\mathrm{even}$&0.98&0.63&0.36\\
   $\text E_\mathrm{min}$&0.97&0.67&0.37\\
   $\text E_\mathrm{min+\textbf{s}^2}$&1.01&0.60&0.36\\\\
   $\Delta^{(3)}_{Z}$& SkI3 & SLy6 & SV-bas\\  \hline
   $\text E_\mathrm{even}$&0.79&0.69&0.39\\
   $\text E_\mathrm{min}$&0.82&0.73&0.40\\
   $\text E_\mathrm{min+\textbf{s}^2}$&0.82&0.66&0.35
  \end{tabular}
 \end{center}
\end{table}

\subsection{Relation between OES and spectral gap}

The OES (\ref{D3N})-(\ref{D3Z}) are closely related to the
pairing gaps which appear in the two-quasiparticle excitation
spectra of the BCS or HFB calculations \cite{Ring_Schuck_book}.
Such spectral gap can be defined  as the average \cite{Bender_EPJA_00}
\begin{equation}
  \overline{\Delta}_q
  =
  \frac{\sum_{k\in q} |u_k v_k \Delta_{k}|}
       {\sum_{k\in q} |u_k v_k|}
  \quad .
\label{eq:spec-gap}
\end{equation}
We use for this purpose $\Delta_{k}=\int d\textbf{r}\phi_k^*(\textbf{r})\Delta_q \phi_k(\textbf{r})$ 
as the state-dependent gap in even and odd nuclei. The weight
$u_\alpha v_\alpha$ concentrates in the pairing-active
region near the Fermi energy \cite{Bender_EPJA_00}. It is the spectral gap in
even nuclei which is used to calibrate the pairing strength by comparison with
the experimental $\Delta^{(3)}_q$ \cite{svbas}. A fully consistent computation
of odd nuclei allows to check the resemblance of the
spectral gap and $\Delta^{(3)}_q$.

\begin{figure}[t]
\includegraphics[width=0.9\linewidth]{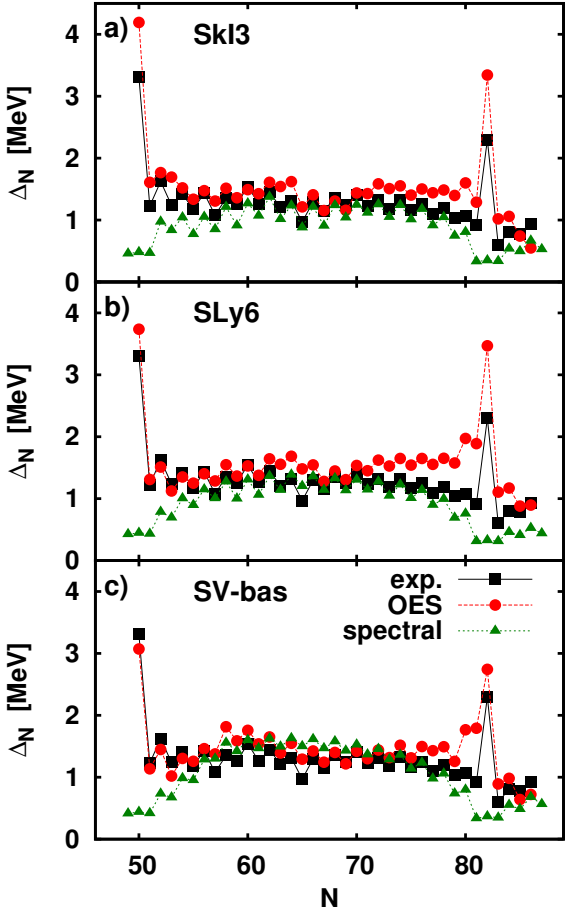}
\caption{
\label{fig:gap-Sn-compare}
Neutron gaps in  Sn isotopes for
three different Skyrme forces.
Compared are the experimental gap from OES
$\Delta^{(3)}_{N, exp}$, the
theoretical gap from OES $\Delta^{(3)}_{N, min+\textbf{s}^2}$,
and the spectral gap as defined in eq. (\ref{eq:spec-gap}).
}
\end{figure}
Fig. \ref{fig:gap-Sn-compare} shows the spectral neutron gap $\overline{\Delta}_N$
together with the theoretical OES $\Delta^{(3)}_{N,\mathrm{min}+\textbf{s}^2}$ and
experimental $\Delta^{(3)}_N$ in Sn isotopes.  There is
an acceptable agreement between $\overline{\Delta}_N$ and
$\Delta^{(3)}_{N,\mathrm{min}+\textbf{s}^2}$ in the mid-shell region but a
large deviation in trend and magnitude near shell closures where pairing
shrinks and shell effects dominate. The magnitude of deviation depends on the
parameterization, which confirms its shell origin. The
deviations in the mid-shell region stay in a range of about 10\% and justify
the resemblance of $\overline{\Delta}_N$ and
$\Delta^{(3)}_{N,\mathrm{min}+\textbf{s}^2}$ as a first guess. The
theoretical results reproduce nicely the experimental data in the pairing-dominated mid-shell regions,
which is not surprising since just these regions were used for the fit of the pairing strength.

\begin{figure}[t]
\includegraphics[width=0.9\linewidth]{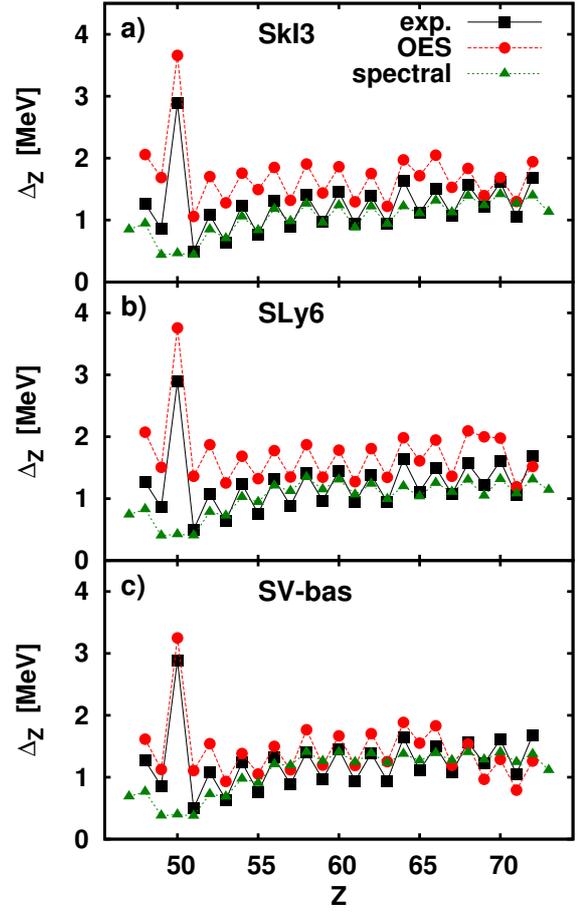}
\caption{
\label{fig:gap-N82-compare}
The same as figure \ref{fig:gap-Sn-compare}
for proton gaps in the isotonic chain N=82.
}
\end{figure}
Fig. \ref{fig:gap-N82-compare} gives the similar comparison for the proton gaps in
N=82 isotones. There is again a huge mismatch between
$\overline{\Delta}_Z$ and $\Delta^{(3)}_{Z,\mathrm{min}+\textbf{s}^2}$
near the Z=50 shell closure. However the agreement in the mid shell region
is not as good as for the neutron gaps and strongly
depends on the Skyrme parameterization. It is thus likely that shell
effects continue to contribute even far off the shell closure. Note that
the deviation is smallest for SV-bas.

It is interesting to note that one sees clearly that the spectral gap
$\overline{\Delta}$ was fitted to the experimental OES in even nuclei
which shows still good enough agreement even in this case having 
generally large deviations.

\begin{figure}[t]
\includegraphics[width=0.9\linewidth]{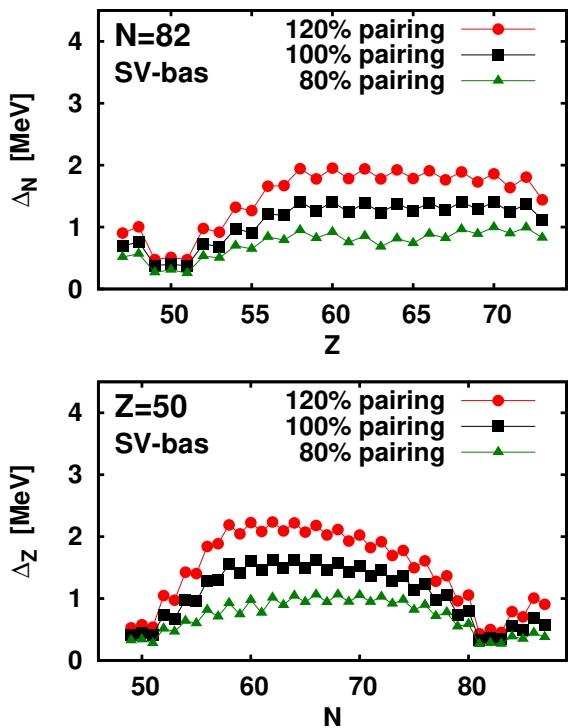}
\caption{
\label{fig:varypair-SVbas}
Theoretical OES gaps for varied pairing strength.
Upper:  $\Delta^{(3)}_{Z, min+\textbf{s}^2}$ in the
isotones with N=82.
Lower:  $\Delta^{(3)}_{N, min+\textbf{s}^2}$ in the Sn isotopes.
}
\end{figure}

The sensitivity of OES to the pairing strength is tested in figure
\ref{fig:varypair-SVbas}. The result corroborates the finding of
the previous two figures. The pairing does not affect the
regions near shell closures but dominates mid shell.
The changes in the gaps are rather constant in mid shell and
exceed those in the pairing strengths (30\% versus 20\%).
A similar amplification was previously found in \cite{Rut99c}.

\subsection{Separation energies}

The neutron and proton separation energies (SE)
\begin{eqnarray}\label{S_N}
  S_N (Z,N) &=& E(Z,N)-E(Z,N-1) \; ,
\\
 S_Z (Z,N) &=& E(Z,N)-E(Z-1,N)
\label{S_NZ}
\end{eqnarray}
represent an important nuclear characteristics. They are sensitive
to all aspects of the nuclear mean field and pairing. Since SE deal
with odd or/and odd-odd nuclei, they can be influenced by TOMF.

\begin{figure}[t]
\includegraphics[width=\linewidth]{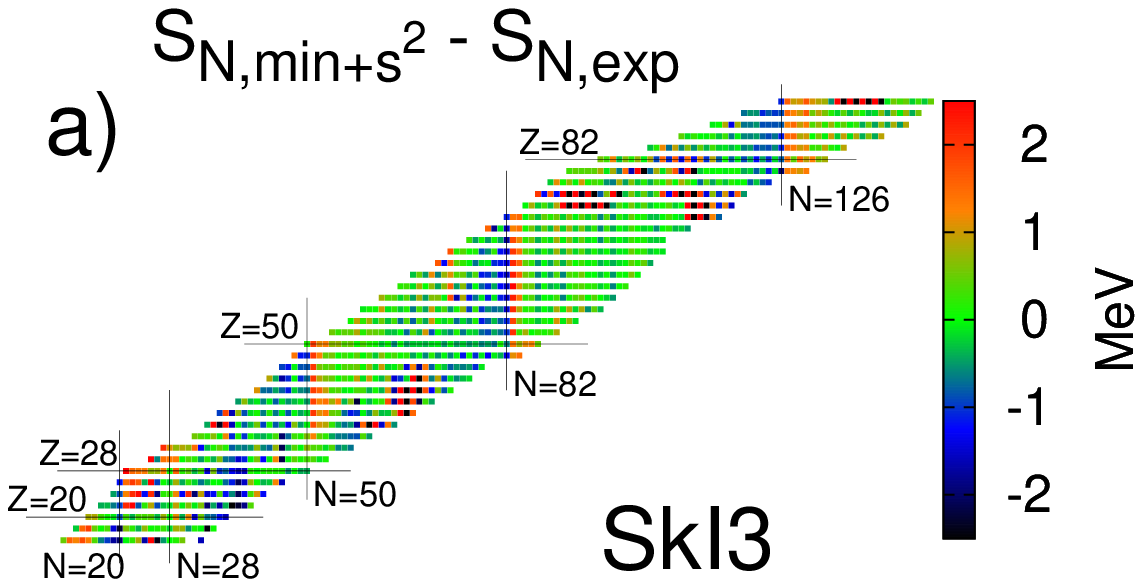}
\includegraphics[width=\linewidth]{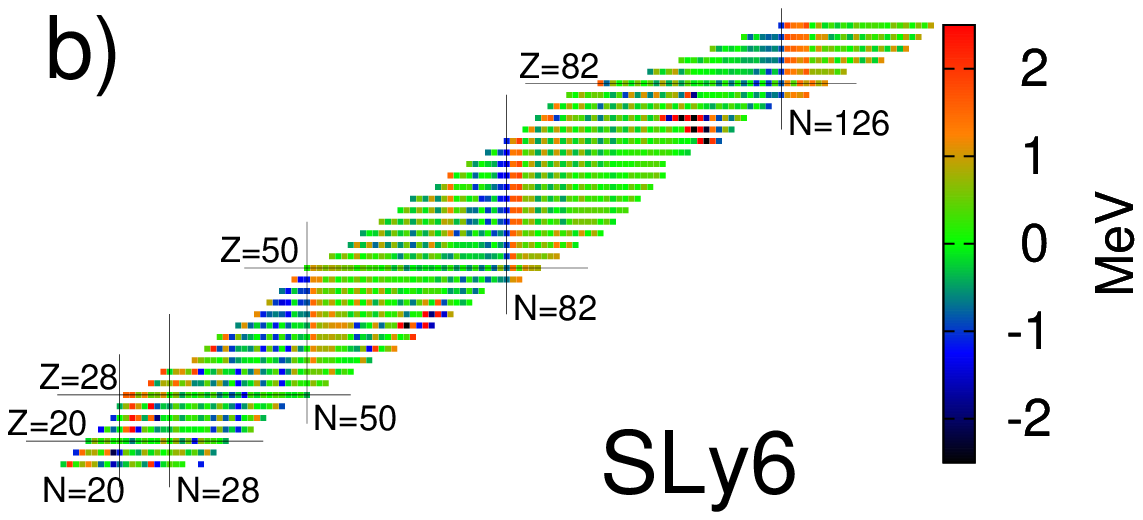}
\includegraphics[width=\linewidth]{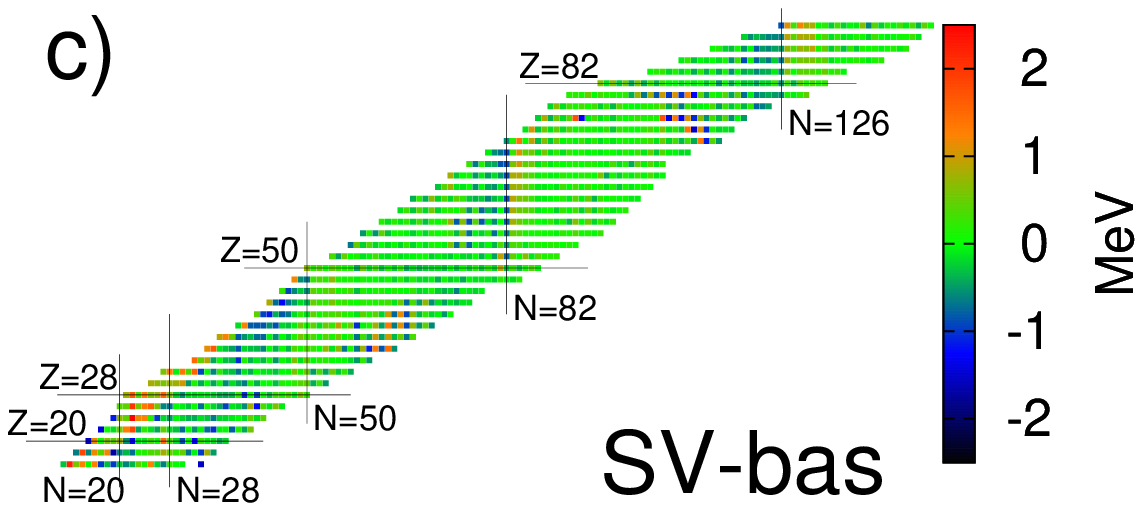}
\caption{\label{fig12}
The systematics of
difference between SHF and experimental \cite{exp_BE}
neutron separation energies,
$S_\mathrm{N, min+\textbf{s}^2} - S_\mathrm{N, exp}$,
drawn in the $N$-$Z$-plane
for the parameterizations SkI3, SLy6, and SV-bas.}
\end{figure}
\begin{figure}[t]
\includegraphics[width=\linewidth]{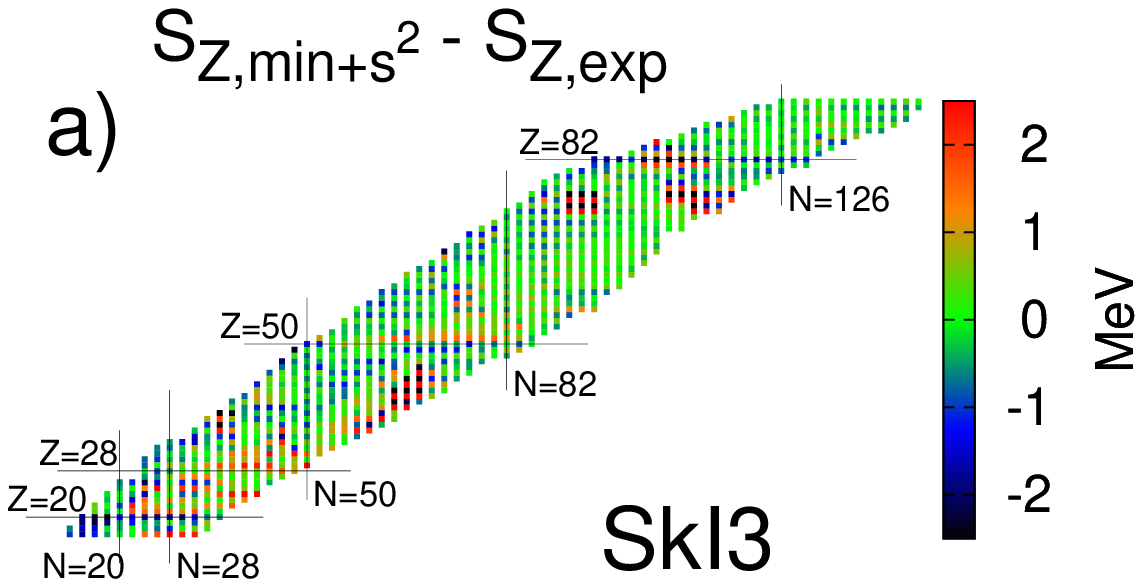}
\includegraphics[width=\linewidth]{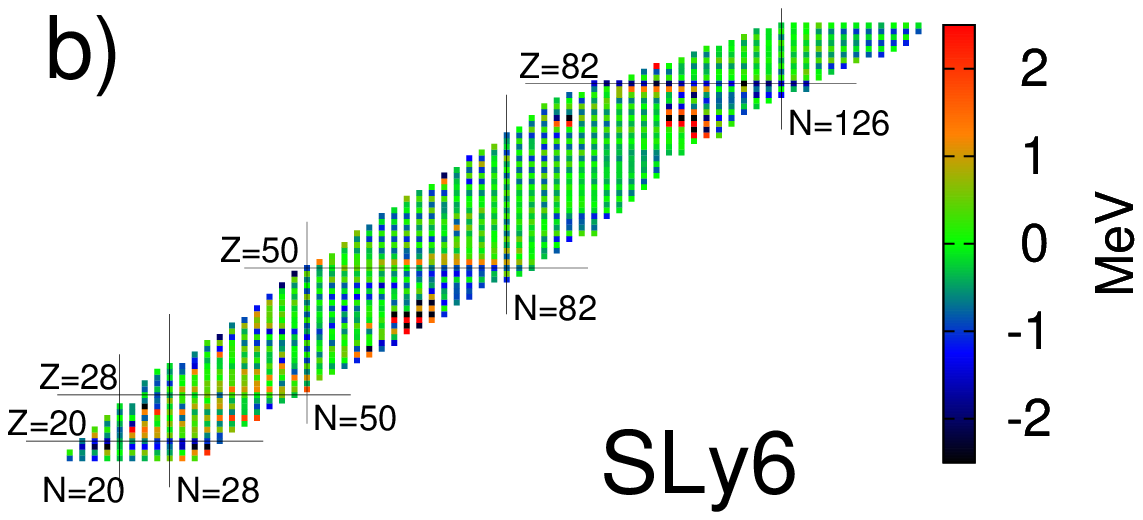}
\includegraphics[width=\linewidth]{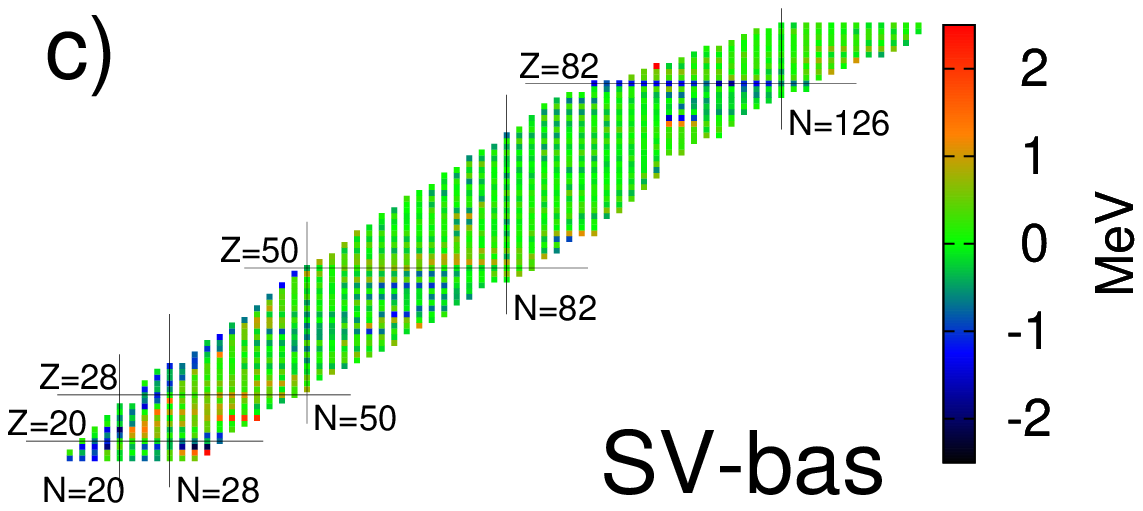}
\caption{\label{fig13}
The same as in Fig. \ref{fig12} for the proton separation energies
$S_\mathrm{Z}$.}
\end{figure}

Figs. \ref{fig12} and \ref{fig13} show the systematics of the
deviations of SHF neutron and proton SE from the experimental
data \cite{exp_BE}. There are pronounced errors along isotonic chains for $S_\mathrm{N}$
and isotopic chains for $S_\mathrm{Z}$. They sensitively depend on the
parameterization and are minimal for SV-bas. Obviously this is a shell effect.
SE show a large jump at shell closures and size of the jump is closely
related to the OES magnitude near the closure.
The OES in turn, depends sensitively on the
effective mass of the underlying model. The differences near the
shell closures in Figs.  \ref{fig12} and \ref{fig13} are thus related to
the effective masses of SHF parameterizations as well. The good
performance of SV-bas indicates once more that its effective mass of
0.9 is favorable.
The regions of well deformed nuclei (remote from the magic Z and N lines)
show generally smaller deviations for all parameterizations.  Instead the
transitional regions (close to the magic Z and N lines) exhibit larger errors.
The latter is probably caused by missing the correlations which are known to
be most strong for transitional nuclei.

\begin{table}[htb]
\caption{\label{tab:error-SnSz}
R.m.s. deviations (in MeV) between SHF and experimental data \cite{exp_BE}
for separation energies $S_{N,Z}$ from Figs. \ref{fig12}-\ref{fig13}.
}
 \begin{center}
  \begin{tabular}{lccc}
   $S_{N}$&SkI3&SLy6 &SV-bas\\  \hline
   $\text E_\mathrm{even}$&1.15&0.83&0.50\\
   $\text E_\mathrm{min}$&1.16&0.85&0.50\\
   $\text E_\mathrm{min+\textbf{s}^2}$&0.80&0.60&0.50\\\\
   $S_{Z}$&SkI3&SLy6 &SV-bas\\  \hline
   $\text E_\mathrm{even}$&0.96&0.83&0.53\\
   $\text E_\mathrm{min}$&0.98&0.87&0.54\\
   $\text E_\mathrm{min+\textbf{s}^2}$&0.97&0.82&0.50
  \end{tabular}
 \end{center}
\end{table}

Table \ref{tab:error-SnSz} summarizes the r.m.s. deviations for SE.
A large difference in performance for the three parameterizations
is seen, where SV-bas shows the best results. The difference
between various options of the energy functional is smaller. In particular,
$\text E_\mathrm{min + \textbf{s}^2}$ gives a slight
improvement of the overall quality.   Figs. \ref{fig12}-\ref{fig13} and
Table \ref{tab:error-SnSz}  altogether indicate that the
choice of the proper Skyrme parameterization
is crucial to achieve a good quality of the SE description
in all regions of the nuclear chart. Transitional nuclei may require
the correlation effects.

\subsection{Excitation spectra of odd nuclei}

The excitation spectrum of odd nuclei is closely related to the
single-particle spectrum of the even-even neighbor.  In a simple
picture one even may deduce the single-particle spectra of the
even-even nucleus from the excitations of the nearby odd systems. We
will now check in more detail the effect of TOMF on the excitation
energies of odd nuclei and the performance of the simple estimate in
terms of the single-particle spectra of the even-even reference
system. The comparison with the experimental data will be also
done. Here, however, one has to keep in mind that the low-energy
excitation spectra of odd nuclei may be heavily influenced by
admixture of vibrational states of the even-even core, resulting in in
a general essential compression of the low energy spectrum.
\cite{Solov_book,Gareev_NP_71}.
\begin{figure}[t]
\includegraphics[width=\linewidth]{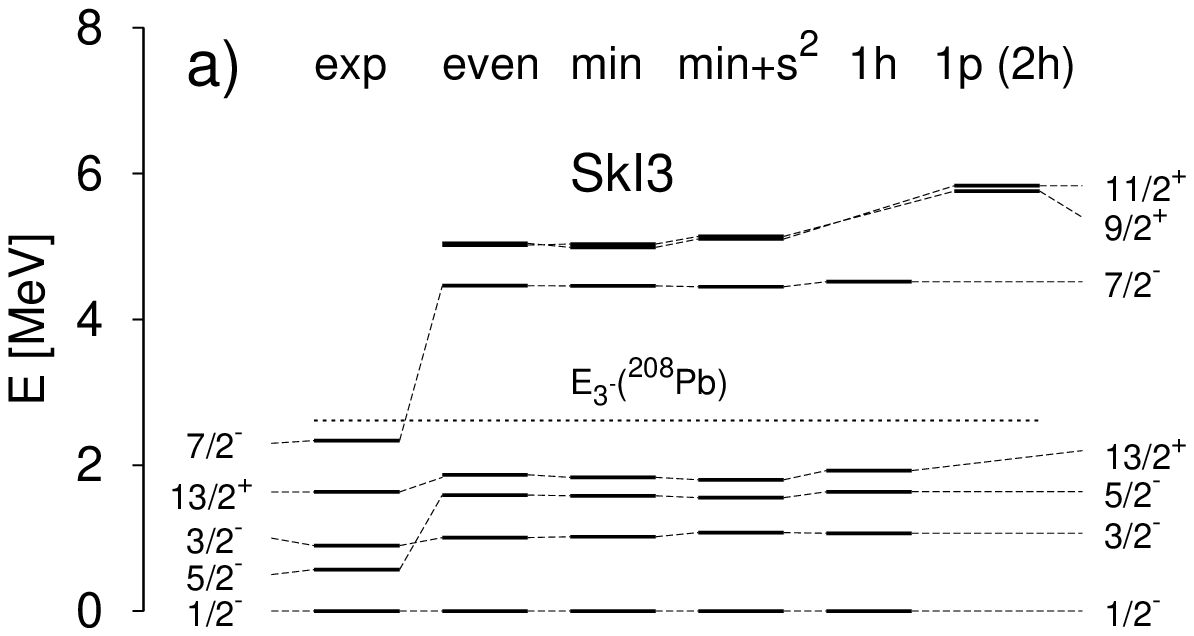}
\includegraphics[width=\linewidth]{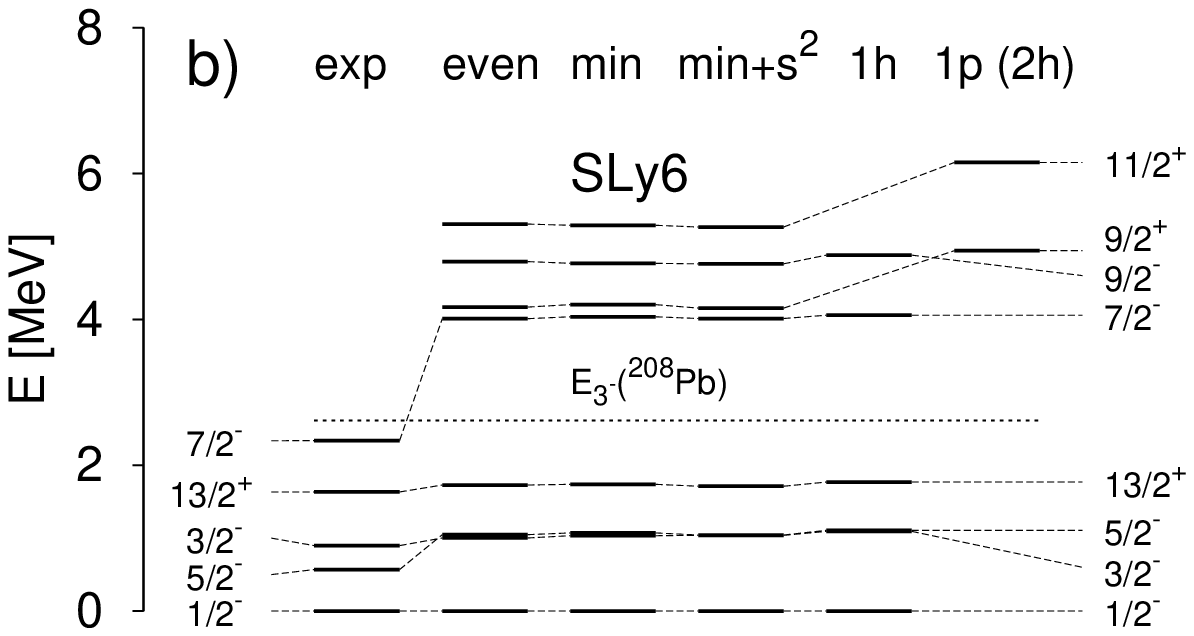}
\includegraphics[width=\linewidth]{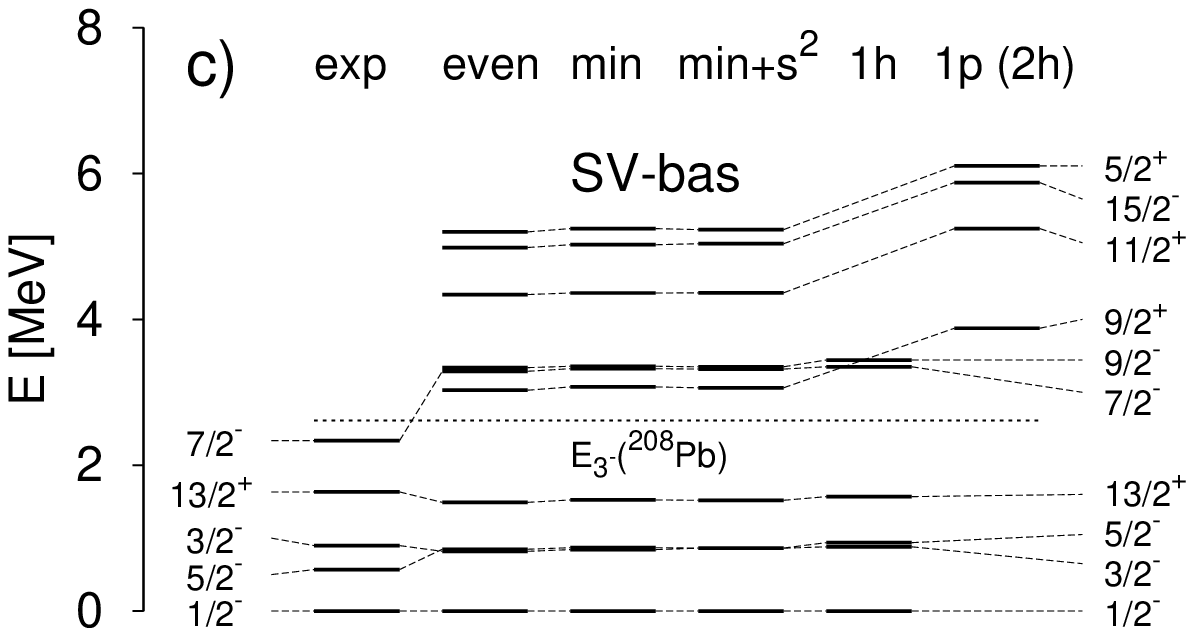}
\caption{\label{fig14}
Excitation neutron spectra in $^{207}$Pb for the parameterizations SkI3 (a),
SLy6 (b), and SV-bas (c). In every panel, the experimental data
\protect\cite{exp_spe_207Pb} are compared
with the SHF results for the options (\ref{eq:op1})-(\ref{eq:op3}).
Also the simple one-hole (1h)
and one-particle-two-hole (1p2h) estimates from the
single-neutron spectrum of $^{208}$Pb are exhibited.
The dotted horizontal line marks energy of the lowest
collective state $3^-$ in $^{208}$Pb. For better view, results for one
and the same level are connected by dash lines. See text for more details.}
\end{figure}

Figure \ref{fig14} shows the excitation spectra in $^{207}$Pb. This
nucleus has a neutron hole with respect to $^{208}$Pb. The simple
excitation model is then deduced from the neutron one-hole (1h) and
one-particle-two-hole (1p2h) excitation spectra in $^{208}$Pb,
as indicated in the figure. In this model, the low-energy spectrum
is dominated by 1h excitations but at higher energies the 1p2h excitations
become energetically competitive.
Such prescription is widely used in various
approaches, see e.g. \cite{Ring_Schuck_book,Solov_book,Gareev_NP_71},
and it obviously does not include the TOMF effects.

These effects are illustrated in Fig. \ref{fig14} for the options
$even$, $mean$, and $min+s^2$ where the TOMF are added step by step.
Unlike the simple model discussed above, these results are consistent
in the sense that for every state the SHF problem is solved independently
by minimization of the total energy. As seen from the figure, the TOMF
impact is negligible. At the same time we have strong rearrangement effect
for 1p2h states, which is only accounted for the self-consistent mean-field
calculations. This makes the simple estimations less reliable, although they
still yield a nice first guess.

Comparison with the experimental spectra shows that the calculated levels
below 2 MeV follow the right sequence and rather well reproduce the
experimental energies, especially for SLy6 and SV-bas. Discrepancies start
for the $7/2^-$ state which has too high energy in all
predictions. Most probably this is caused by omitting the vibrational
admixture which generally lower and compress the low-energy spectrum.
Figure \ref{fig14} shows the energy of the lowest collective
state $3^-$ in the even-even core $^{208}$Pb, which can serve as
a rough energy indicator of the onset of core vibrational admixture.
(A more complete estimate would, of course, require to know also
the coupling matrix elements.)

\begin{figure}[t]
\includegraphics[width=\linewidth]{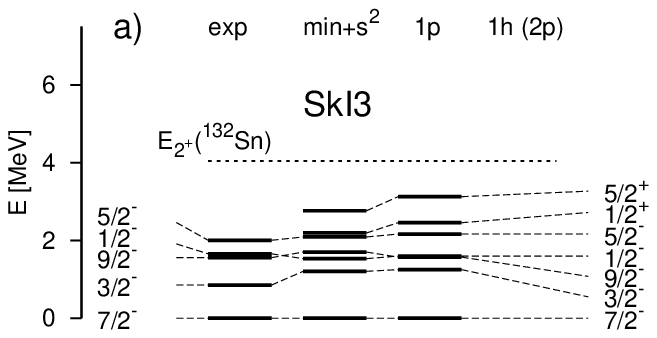}
\includegraphics[width=\linewidth]{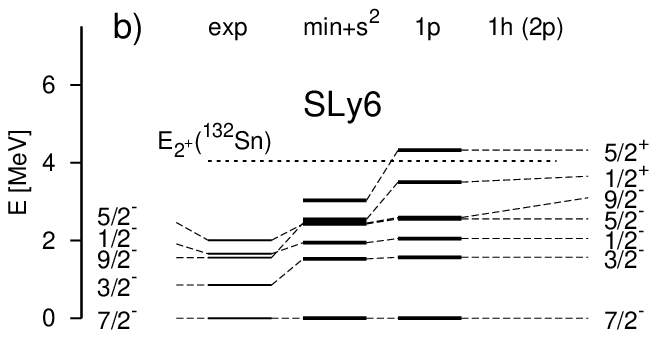}
\includegraphics[width=\linewidth]{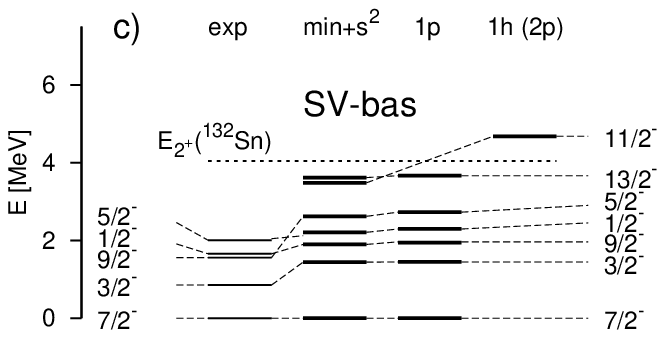}
\caption{\label{fig15}
The same as in Fig. \ref{fig14} but for the neutron spectrum in $^{133}$Sn
and the option $min+s^2$ alone. The dotted line marks the lowest
collective state $2^+$ in reference even-even core $^{132}$Sn.
The experimental data are taken from \protect\cite{exp_spe_133Sn}.
}
\end{figure}

Results of similar calculations for neutron-excess nucleus $^{133}$Sn
are shown in Fig. \ref{fig15}. This is the case where the neutron particle (1p)
states in $^{132}$Sn are explored.  Different TOMF options  are
again found to make a negligible difference. Thus we display here
only results from the final option $min+s^2$. The simple 1p
estimate provides a nice first guess whose reliability, however, fades with
increasing excitation energy, particularly if one approaches the
vicinity of the $2^+$ core excitation.  The comparison
with the experimental energies in $^{133}$Sn is less convincing than in
$^{207}$Pb. All excitation energies are overestimated.
This is not a problem of spectral density because all three parameterizations
yield about the same trends. The most probable reason of the mismatch
could be neglecting the core vibrational admixture whose impact in $^{133}$Sn
might be stronger than in $^{207}$Pb and so affect even the lowest excitations.
The comparison of the simple estimations with the $min+s^2$ case exhibits a strong
rearrangement effect pertinent to self-consistent calculations.

\begin{figure}[t]
\includegraphics[width=\linewidth]{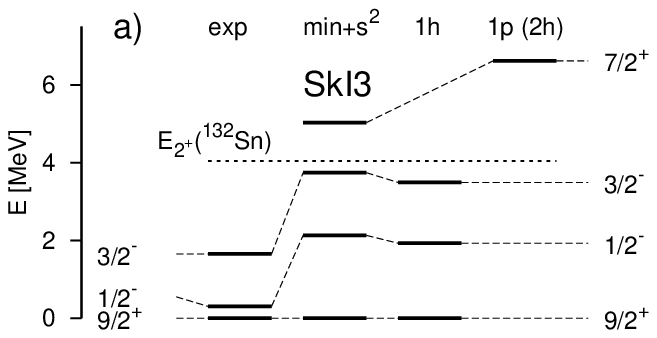}
\includegraphics[width=\linewidth]{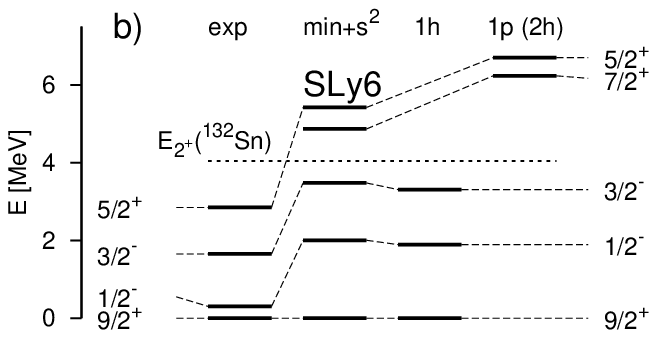}
\includegraphics[width=\linewidth]{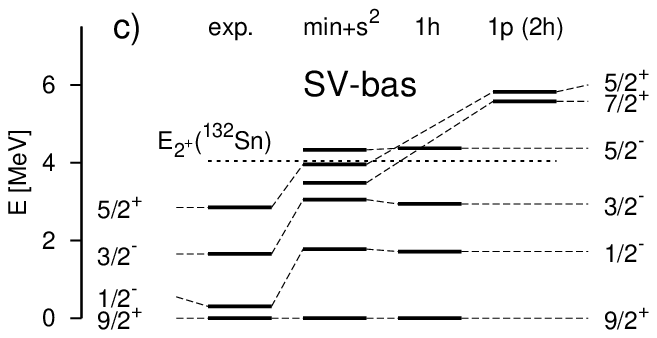}
\caption{\label{fig16}
The same as in Fig. \ref {fig15} for the proton spectrum in  $^{131}$In.
The reference even-even core is $^{132}$Sn. The experimental
data are taken from \protect\cite{exp_spe_131In}. }
\end{figure}

Fig. \ref{fig16} shows the spectra for $^{131}$In, the proton-hole
neighbor of $^{132}$Sn. The general trends are very similar to the previous
figure. Again the simple 1h spectrum provides a good first
guess, there is a strong rearrangement effect, and the calculated energies
overestimate the experimental data. The later problem appears for all neighbors of
$^{132}$Sn while the neighbors of $^{208}$Pb perform fairly well.
This systematic deviation is still awaiting an explanation.

\section{Conclusions}
\label{sec_concl}

In this paper, we have discussed the description of odd-even and even-odd
nuclei in the framework of the Skyrme-Hartree-Fock (SHF) method with BCS
pairing.
Particular attention was paid to the influence of the time-odd mean
fields (TOMF) in the Skyrme functional which are brought into action by the
breaking of time-reversal symmetry through the odd nucleon. Three options were
compared: $even$ with time-even terms only and omitted TOMF, $min$ with adding
the minimum of TOMF to restore the Galilean invariance, and $min+s^2$ with
additional spin couplings. Gradient spin couplings were omitted as leading
to instabilities. The odd nuclei were computed within BCS with blocking the
state occupied by the odd nucleon. A bunch of blocked states was
computed self-consistently thus yielding the ground state and first few excitations
in the odd nucleus. The nuclear properties which might be affected by TOMF,
binding energies (BE), odd-even staggering (OES), separation energies (SE)
and low-energy single-particle spectra were systematically investigated
with three Skyrme parameterizations SkI3, SLy6, and SV-bas for the
nuclear chart with $16 \le Z \le 92$.

The TOMF effects are found generally small and for the spectra even
negligible. They affect the results much weaker than the choice of the
Skyrme parameterization or rearrangement in self-consistent calculations.
For BE, the TOMF effect is maximal for light nuclei and
rapidly decreases with increasing nuclear mass number, in accordance to
the previous results \cite{Satula_99,Afa_PRC_10_RMF}. The change in
BE is 100 - 300 keV and TOMF may lead to both less and more binding,
depending on the Skyrme parameterizations. The BE description is best
with the minimal Galilean invariant option $min$ while OES and SE prefer the
full functional option $min+s^2$. The OES can be modified
by $\sim$ 200 keV and SE by $\sim$ 200 keV.
These results are fluctuations and sometimes conflicting since the currents
and spin densities give opposite contributions and partly compensate
each other, thus producing a fragile balance.
Obviously, so weak and unstable TOMF effects cannot be used
for the upgrade of the Skyrme functional in the spin channel.

Some additional important points should be mentioned.
First of all, the primary importance of the proper choice of the Skyrme
parameterization has to be emphasized.  This affects the results
much more than TOMF. For all the nuclear characteristics
considered in the present exploration, the best description was obtained
with the recent parameterization SV-bas. Note that SV-bas
has a large effective mass $m^*/m=0.9$ and its fit involves isotopic 
as well as isotonic chains.

Second, differences of binding energies, like OES and SE, are especially
sensitive to the shell and pairing details. The OES is dominated by pairing in the
mid shell regions and by shell effects near shell closures. Thus a strong
dependence on the SHF parameterization is seen at shell closures and very
little in mid shell regions where the agreement with experimental OES  is generally
satisfying. It ought to be reminded that the pairing strength is calibrated to
the spectral gap which deviates from OES. Hence the relation between these
pairing characteristics was analyzed.

Third, low lying excitation spectra for odd nuclei next to doubly magic ones have 
been analyzed. There are cases which work nicely and others which do not fit
so well. The mismatch is probably caused by the coupling with
vibrational and rotational modes of the even-even core. They are especially
strong in deformed and transitional nuclei. The correlations could considerably
improve the description of BE, OES, and SE, first of all in mid shell nuclei.
Besides, the correlations could be of a crucial importance
for description of the low-energy single-particle spectra of odd nuclei.
Being too dilute in the mean-field picture,
the spectra can become more compressed and closer to the
experimental data.

\begin{acknowledgement}
V.O.N. thanks the support from DFG RE-322/12-1 and
Heisenberg-Landau (Germany - BLTP JINR) grants.
\end{acknowledgement}

\appendix 

\section{Relation between parameters of the functional}
\label{app:params}

The SHF method can be alternatively formulated by starting
since the density-dependent zero-range Skyrme force
\begin{align}
  \hat V_{Skyrme}
  &=\hat V_{12}  + \hat V_{dd}+ \hat V_{LS}
\\
  \hat V_{12}
  &=t_0(1+x_0 \hat{ P}_\sigma)\delta(\textbf r_1-\textbf r_2)
\\
  &\qquad
  + \frac{t_1}{2}(1+x_1\hat{P}_\sigma)\delta(\textbf r_1-\textbf r_2)
  \hat{\boldsymbol k}^2
\nonumber\\
  &\qquad
  + \frac{t_1}{2}(1+x_1\hat{P}_\sigma){\hat{\boldsymbol{k}}}'^{2}
  \delta(\boldsymbol r_1-\boldsymbol r_2)
\nonumber\\
  &\qquad
  + t_2(1+x_2\hat P_\sigma)\hat{\boldsymbol k}
  \delta(\textbf r_1-\textbf r_2)\hat{\boldsymbol k}
\nonumber\\
  \hat V_{dd}
  &= \frac{t_3}{6}(1+x_3\hat P_\sigma)
  \rho^\alpha\left(\frac{\textbf r_1 + \textbf r_2}{2}\right)
  \delta(\textbf r_1 - \textbf r_2)
\\
  \hat V_{LS}
  &=
  it_4(\hat{\boldsymbol{\sigma}} _1 + \hat{\boldsymbol{\sigma}}_2)
  \cdot \hat{\boldsymbol k}' \times \delta(\textbf{r}_1-\textbf{r}_2)
  {\mathbf{k}}
\end{align}
with the spin-exchange operator
$\hat P_\sigma=\frac{1}{2}(1+\hat{\boldsymbol
  \sigma}_1\hat{\boldsymbol{\sigma}}_2)$,
the difference momentum
$\hat{\boldsymbol k} =
-\frac{i}{2 }\left(\stackrel{\rightarrow}{\boldsymbol\nabla}_1
  -\stackrel{\rightarrow}{\boldsymbol\nabla}_2\right)$
acting to the right, and its counterpart
$\hat{\boldsymbol
  k}'=\frac{i}{2 }\left(\stackrel{\leftarrow}{\boldsymbol\nabla}_1
      -\stackrel{\leftarrow}{\boldsymbol\nabla}_2\right)$ acting to the left.
Then the  SHF functional (\ref{eq:Sk-even}-\ref{eq:skyrme_funct})  is
obtained as the expectation value of this force with a Slater state.
This yields one-to-one correspondence of the force parameters
$t_i, x_i$ with the parameters $b_i, b_i',
\tilde{b}_i, \tilde{b}'_i$ from (\ref{eq:Sk-even}-\ref{eq:skyrme_funct}):
\begin{equation}
  \begin{array}{rclcrcll}
   b_0
   &=&
   t_0(1+\frac 12 x_0)
   &&
   b'_0
   &=&
   t_0(\frac 12 + x_0)
   &
  \\
   b_1
   &=&
   \frac {t_1}{4}(1+\frac 12 x_1)
   &&
   b'_1
   &=&
   \frac {t_1}{4}(\frac 12 + x_1)
   &
  \\

   &&
   +\ \frac{t_2}{4}(1+\frac 12 x_2)
   &&

   &&
   -\ \frac{t_2}{4}(\frac 12 + x_2)
   &
  \\
   b_2
   &=&
   \frac {3t_1}{8}(1+\frac 12 x_1)
   &&
   b'_2
   &=&
   \frac{3t_1}{8}(\frac 12 +x_1) 
   &
  \\

   &&
   -\ \frac{t_2}{8}(1+\frac 12x_2)
   &&

   &&
   +\ \frac{t_2}{8}(\frac 12 + x_2)
   &
  \\
   b_3
   &=&
   \frac 14t_3(1+\frac 12 x_3)
   &&
   b'_3
   &=&
   \frac 14t_3(\frac 12 +x_3)
   &
  \\
   b_4
   &=&
    \frac 12t_4
   &&
   b'_4
   &=&
    \frac 12 t_4
   &
   \\
   \tilde{b}_0
   &=&
    \frac 14t_2x_0
   &&
   \tilde{b}'_0
   &=&
    \frac 14 t_0
   &
   \\
   \tilde{b}_1
   &=&
    \frac 18(t_1x_1+t_2x_2)
   &&
   \tilde{b}'_1
   &=&
    \frac 18(t_1-t_2)
   &
   \\
   \tilde{b}_2
   &=&
    \frac{1}{16}(3t_1x_1-t_2x_2)
   &&
   \tilde{b}'_2
   &=&
    \frac {1}{16}(3t_1+t_2)
   &
  \\
   \tilde{b}_3
   &=&
    \frac{1}{24}t_3x_3
   &&
   \tilde{b}'_3
   &=&
    \frac {1}{24}t_3
   &.
  \end{array}
\end{equation}
\section{Single-particle wave function}
\label{app:spwf}

The single-particle wave function in the
cylindrical coordinates is the spinor
\begin{equation}\label{eq:psi_i_spin}
\phi_{k}(\textbf{r}) =
\left(\begin{array}{c} {\phi_{k}^{(+)}(\textbf{r})} \\
{\phi_{k}^{(-)}(\textbf{r})} \end{array}
 \right)
=
\left(\begin{array}{c} {R_{k}^{(+)}(\rho,z) e^{im_{k}^{(+)}\vartheta}} \\
{R_{k}^{(-)}(\rho,z) e^{im_{k}^{(-)}\vartheta}} \end{array}
 \right) ,
\end{equation}
with
\begin{equation}\label{eq:new_m}
m_{k}^{(s)} = K_k - \frac{1}{2} s \;,
 \quad
m_{k}^{(-s)} = K_k + \frac{1}{2} s
\end{equation}
and $K_k$ is the projection of the total momentum of k-state
on the symmetry z-axis.

\end{document}